\documentclass[reprint,superscriptaddress,amsmath,amssymb,aps,prb,floatfix]{revtex4-1}

\usepackage{graphicx}
\usepackage{multirow}
\usepackage{bm}
\usepackage{hyperref}
\usepackage{todonotes}
\usepackage{nicefrac}
\usepackage{mathtools}
\usepackage[normalem]{ulem}
\usepackage[version=4]{mhchem}

\definecolor{fj_color}{cmyk}{1, 0.3, 0, 0}
\definecolor{cwh_color}{cmyk}{0, 0.8, 0.8, 0}
\definecolor{ys_color}{cmyk}{0.5, 0.8, 0.0,0}

\newcommand{\SRO}{Sr\textsubscript{2}RuO\textsubscript{4}}

\begin{document}
\title{\boldmath{$T_c$} and the elastocaloric effect of \SRO{} under \boldmath{$\langle 110 \rangle$} uniaxial stress: no indications of transition splitting}

\author{Fabian Jerzembeck} 
\email{Fabian.Jerzembeck@cpfs.mpg.de}
\affiliation{Max Planck Institute for Chemical Physics of Solids, D-01187 Dresden, Germany}

\author{You-Sheng Li}
\affiliation{Max Planck Institute for Chemical Physics of Solids, D-01187 Dresden, Germany}
\affiliation{Department of Physics, National Taiwan University, Taipei, 10617, Taiwan}

\author{Grgur Palle}
\affiliation{Institute for Theory of Condensed Matter, Karlsruhe Institute of Technology, 76131 Karlsruhe, Germany }

\author{Zhenhai Hu}
\affiliation{Max Planck Institute for Chemical Physics of Solids, D-01187 Dresden, Germany}

\author{Mehdi Biderang}
\affiliation{Department of Physics, University of Toronto, 60 St. George Street, Toronto, Ontario, M5S 1A7, Canada}

\author{Naoki Kikugawa}
\affiliation{National Institute for Materials Science, Tsukuba 305-0003, Japan}

\author{Dmitry A. Sokolov}
\affiliation{Max Planck Institute for Chemical Physics of Solids, D-01187 Dresden, Germany}

\author{Sayak Ghosh}
\affiliation{Laboratory of Atomic and Solid State Physics, Cornell University, Ithaca, New York 14853, USA}

\author{Brad J. Ramshaw}
\affiliation{Laboratory of Atomic and Solid State Physics, Cornell University, Ithaca, New York 14853, USA}
\affiliation{Canadian Institute for Advanced Research, Toronto, Ontario, Canada}

\author{Thomas Scaffidi}
\affiliation{Department of Physics and Astronomy, University of California, Irvine, California 92697, USA}
\affiliation{Department of Physics, University of Toronto, 60 St. George Street, Toronto, Ontario, M5S 1A7, Canada}

\author{Michael Nicklas}
\affiliation{Max Planck Institute for Chemical Physics of Solids, D-01187 Dresden, Germany}

\author{J\"{o}rg Schmalian}
\affiliation{Institute for Theory of Condensed Matter, Karlsruhe Institute of Technology, 76131 Karlsruhe, Germany }
\affiliation{Institute for Quantum Materials and Technologies, Karlsruhe Institute of Technology, 76131 Karlsruhe, Germany }

\author{Andrew P. Mackenzie}
\email{Andy.Mackenzie@cpfs.mpg.de}
\affiliation{Max Planck Institute for Chemical Physics of Solids, D-01187 Dresden, Germany}
\affiliation{Scottish Universities Physics Alliance (SUPA), School of Physics and Astronomy, University of St. Andrews, St. Andrews KY16 9SS, United Kingdom}

\author{Clifford W. Hicks}
\email{C.Hicks.1@bham.ac.uk}
\affiliation{Max Planck Institute for Chemical Physics of Solids, D-01187 Dresden, Germany}
\affiliation{School of Physics and Astronomy, University of Birmingham, Birmingham B15 2TT, United Kingdom}

\begin{abstract}
There is considerable evidence that the superconductivity of \SRO{} has two components. Among this evidence is a jump in the shear elastic modulus $c_{66}$ at the critical temperature
$T_c$, observed in 
ultrasound measurements.
Such a jump is forbidden for homogeneous single-component order parameters, and implies that $T_c$ should develop as a cusp under the application of shear strain with $\langle 110 \rangle$ principal axes. 
 This shear strain should split the onset temperatures of the two components, if they coexist, or select one component if they do not.  
Here, we report measurements of $T_c$ and the elastocaloric effect of \SRO{} under uniaxial stress applied along the $[110]$ lattice direction. Within experimental resolution,
we resolve neither a cusp in the stress dependence of $T_c$, nor any second transition in the elastocaloric effect data.  We show that reconciling these null results with
the observed jumps in $c_{66}$ requires extraordinarily fine tuning to a triple point of the Ginzburg-Landau parameter space. In addition, our results are inconsistent with
homogeneous time reversal symmetry breaking at a temperature $T_2 \leq T_c$ as identified in muon spin relaxation experiments.  
\end{abstract}

\maketitle

Although it has a critical temperature $T_c$ of only 1.5~K, \SRO{} has become one of the most-studied unconventional superconductors. 
This is in part because even though the normal state of \SRO{} is extraordinarily well-characterised, the pairing mechanism and superconducting order parameter remain unclear~\cite{Maeno94_Nature, Mackenzie03_RMP,Maeno12_JPSJ, Kallin12_RepProgPhys, Mackenzie17_npj, Maeno24_JPSJ}. Given the extremely high purity of the crystals available for experimental investigation~\cite{Mackenzie98_PRL}, this should be a soluble
problem, and it has become a benchmark for the progress of the broader field of unconventional superconductivity.

As often happens when a large number of experiments are performed on a single material, the results and/or interpretations of some experiments disagree. 
While it is appropriate for theory to attempt to reconcile apparently contradictory results, the possibility of experimental error must also be kept in mind. 
It can be subtle. In the history of \SRO{}, a conflict existed for nearly two decades between two different probes of the competition between superconducting condensation energy and magnetic polarisation energy. 
Pauli critical
field limiting~\cite{Yonezawa13_PRL, Kinjo22_Science} was consistent with even-parity spin-singlet order, but the magnetic polarizability of the superconducting state measured by the NMR Knight shift~\cite{Ishida98_Nature} contradicted that conclusion, leading to extensive discussion of spin-triplet order parameters. 
The issue was resolved only after a
systematic error in the original NMR measurements was uncovered~\cite{Pustogow19_Nature}.
Although some researchers continue to explore the possibility of spin-triplet pairing in
\SRO{}~\cite{Cai22_PRB, Huang21_PRR, Yasui20_npj, Gupta20_PRB}, the weight of recent evidence is now strongly in favour of spin-singlet, even-parity order~\cite{Pustogow19_Nature, Steppke17_Science, Ishida20_JPSJ, Petsch20_PRL, Chronister21_PNAS, Jerzembeck23_PRB}. 

This experience provides strong motivation to check other apparently settled experimental facts about the superconductivity of \SRO{}. 
A major question is whether the
superconducting order parameter breaks time reversal symmetry. 
It has long been widely accepted as fact that it does, on the basis of muon spin rotation ($\mu$SR), Kerr rotation, and
Josephson junction data~\cite{Xia06_PRL, Luke98_Nature, Grinenko21_NatPhys, Grinenko21_NatComm, Kidwingira06_Science, Anwar13_SciRep}. 
However, some expected experimental
signatures have not been observed~\cite{Kittaka09_PRB, Hicks10_PRB}.

Recently, it has become possible to test for two expected consequences of time reversal symmetry breaking (TRSB) in \SRO{} under uniaxial pressure. 
The uniaxial pressure should
break the degeneracy of the order parameter components required to produce a TRSB state, yielding, firstly, a cusp in the stress dependence of $T_c$ centered on zero
pressure~\cite{SigristUeda91, Walker02_PRB}, and, secondly, a splitting of the transition under nonzero pressure that should be observable in thermodynamic data. 
Under uniaxial pressure along the [100] lattice direction, neither effect has been observed, in spite of several searches~\cite{Hicks14_Science, Watson18_PRB, Li21_PNAS}. However, transition splitting was observed in $\mu$SR measurements, a non-thermodynamic probe, under [100] uniaxial stress~\cite{Grinenko21_NatPhys}.
One possible interpretation of this discrepancy is that the thermodynamic measurements of Refs.~\cite{Hicks14_Science, Watson18_PRB, Li21_PNAS} were not sensitive enough to detect the second transition.

There is therefore a premium on extending thermodynamic studies to a situation in which there is more guidance on expected thermodynamic quantities.
Recent observations of a jump in the elastic modulus $c_{66}$ at $T_c$, determined via ultrasound measurements~\cite{Ghosh21_NatPhys, Benhabib21_NatPhys}, provide such guidance for uniaxial stress applied along the $[110]$ lattice direction. 
This stress axis has largely been neglected because the coupling of the electronic structure of \SRO{} to stress applied along the $[110]$ direction is weak~\cite{Hicks14_Science, Lupien01_PRL}.  
However, the observed magnitudes of the jumps in $c_{66}$ imply, through Ehrenfest relations that we derive below, that the cusp and splitting should, surprisingly, be easily observable under stress along this direction.

We report results of high-resolution studies of both the magnetic susceptibility and elastocaloric effect under [110] uniaxial pressure.
Within tight limits, we resolve neither a cusp nor transition splitting.
We show that our results cannot be plausibly reconciled with the observed jumps in $c_{66}$ under assumption of a homogeneous superconducting state--- the level of tuning implied is implausibly fine. 

Combining our results with those from previous work on $[001]$ uniaxial pressure allows a prediction for the dependence of $T_c$ on hydrostatic pressure. 
We find good agreement with measurements of $T_c$ under hydrostatic pressure~\cite{Forsythe02_PRL}, which shows that our data are thermodynamically consistent with previous results.
However, our data are not consistent with recently-reported $\mu$SR results, in which a transition splitting under [110] stress was reported~\cite{Grinenko23_PRB}.


While presenting negative results is infrequently done, two reasons motivated our efforts for doing so.
In the context of \SRO{}, we believe that our findings make
a bulk, thermodynamic superconducting state that breaks time reversal symmetry exceedingly unlikely.  
They also call into question the existence of any two-component order parameter in \SRO{}.  
The robustness  of our conclusion stems from the special place held by thermodynamics in understanding the physics of many-body systems. 
Our results therefore narrow down the search for the symmetry of the pairing state of a material that has been  emblematic to the field of unconventional superconductivity.


\section{Strain components}

To frame the discussion in the paper, we introduce a notation for strains.  We will use the symbols $\varepsilon_{110}$ and $\sigma_{110}$ to denote the strain and stress along the $[110]$ lattice direction, under conditions of
uniaxial stress.  
When these symbols are used, it is assumed that there are also transverse strains due to the Poisson effect.
Based on the elastic moduli at 4~K reported in Ref.~\cite{Ghosh21_NatPhys}, $\sigma_{110} = (187\;\text{GPa}) \times \varepsilon_{110}$.

\begin{figure}[ptb]
\includegraphics[width=85mm]{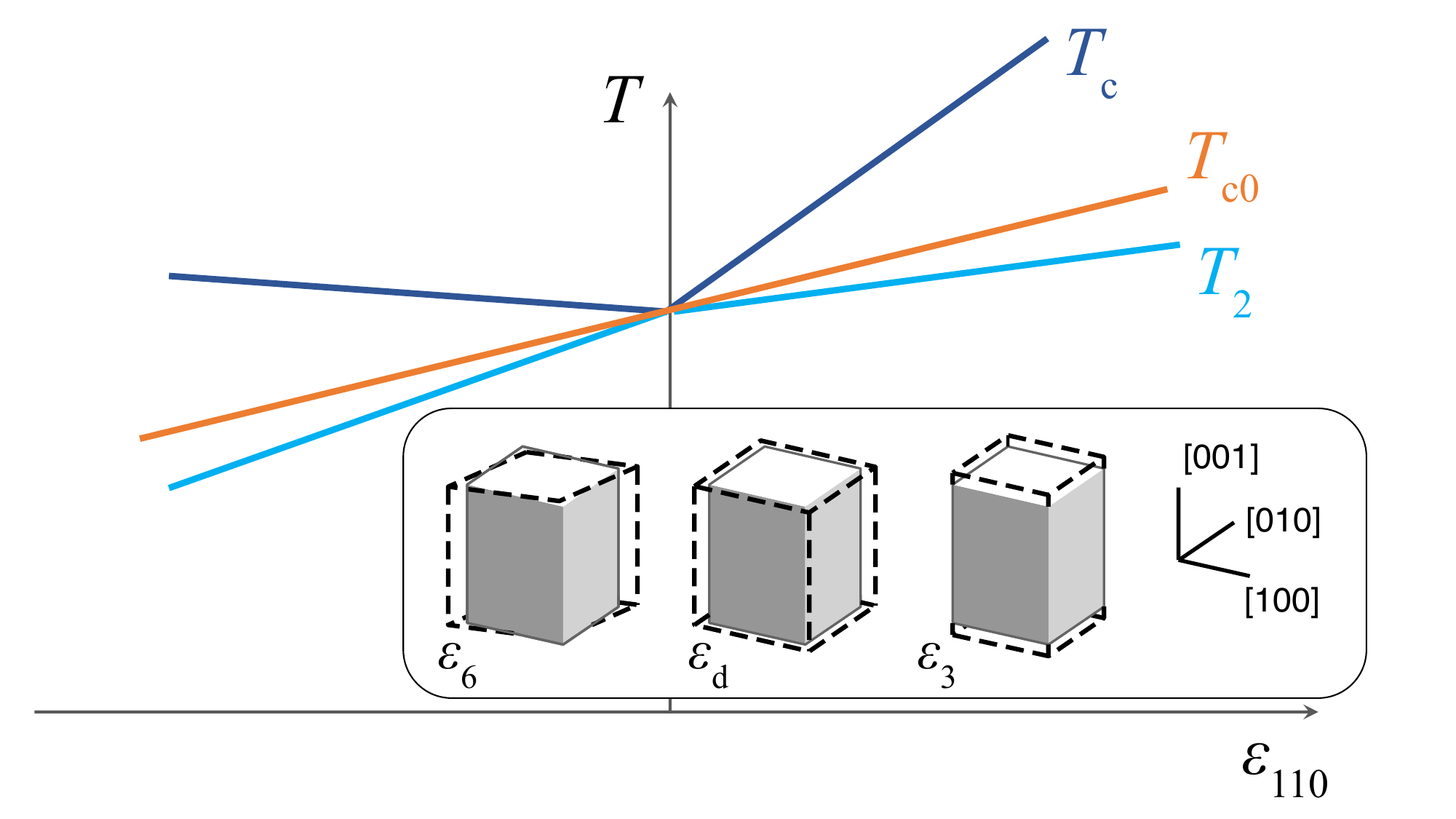}
\caption{Schematic dependence of the phase transition temperatures on strain $\varepsilon_{110}$. Orange line: for single-component and some two-component order parameters, the
onset temperature of superconductivity derives only from the A\textsubscript{1g}-symmetric strain components, $\varepsilon_d$ and $\varepsilon_3$, and so
varies smoothly across $\varepsilon_{110} = 0$.  Dark blue: For $(d_{xz},d_{yz})$ pairing and accidentally degenerate $(s,d_{xy})$ and $(d_{x^2-y^2},g_{xy(x^2-y^2)})$ pairing a
cusp, i.e., a sudden change of slope in $T_c(\varepsilon_{110})$, occurs, due to coupling to the shear strain component $\varepsilon_6$. Light blue: in some cases (see
Fig.~\ref{fig_schematicPhaseDiagrams}) there is a second transition at $T_2$ below $T_c$, also with a cusp. The inset shows the components of the applied strain when uniaxial stress is applied
along a $\langle 110 \rangle$ direction. 
}
\label{fig: Tschematic}
\end{figure}

The strain can be resolved into components.  We choose here to resolve it into shear strain $\varepsilon_6$, $c$-axis strain $\varepsilon_3$, and in-plane dilatation
$\varepsilon_d \equiv \varepsilon_1 + \varepsilon_2$, where $\varepsilon_1$ through $\varepsilon_6$ are the strain components expressed in the standard Voigt notation.
While $\varepsilon_d$ and $\varepsilon_3$ transform under the trivial representation A\textsubscript{1g} of the point group, $\varepsilon_{6}$ transforms under
B\textsubscript{2g}.  
These three strain components are illustrated in the inset of Fig.~\ref{fig: Tschematic}.  
The 4~K elastic moduli from Ref.~\cite{Ghosh21_NatPhys} yield:
\begin{align}
\varepsilon_d &= \alpha_d \,\sigma_{110}, \nonumber \\
\varepsilon_3 &= \alpha_3 \,\sigma_{110}, \nonumber \\
\varepsilon_6 &= \alpha_6 \, \sigma_{110},
\label{eq_strain_components}
\end{align}
with $\alpha_d = 0.00307\;\text{GPa}^{-1}$, $\alpha_3 = -0.00102\;\text{GPa}^{-1}$, and $\alpha_6 = 0.00765\;\text{GPa}^{-1}$.

For all possible order parameters, nonzero $\varepsilon_d$ and $\varepsilon_3$ result in a smooth variation of $T_c$ which is linear in strain to leading order.
We label this line as $T_{c0}$ in Fig.~\ref{fig: Tschematic}.
A leading-order coupling to $\varepsilon_6$ is permitted only for certain two-component order parameters,
and it results in a cusp in the strain dependence of $T_c$, that is, a discontinuity in slope of magnitude $2 |dT_c/d\varepsilon_6|$:
\begin{equation}
\Delta T_c\left(\varepsilon_d,\varepsilon_3,\varepsilon_{6}\right)=\frac{dT_c}{d\varepsilon_d}\varepsilon_d+\frac{dT_c}{d\varepsilon_3}\varepsilon_3+\left|
\frac{dT_c}{d\varepsilon_{6}}\right| \left| \varepsilon_{6}\right|+\cdots,
\end{equation}
where the ellipsis denotes higher-order terms and will be suppressed from now on. 
Below, we show that, for even-parity pairing states, such a cusp occurs for symmetry protected
two-component order parameters that combine $\left(d_{xz},d_{yz}\right)$ Cooper pairs, and for accidentally-degenerate two-component order parameters that combine
$\left(s,d_{xy}\right)$ or $\left(d_{x^{2}-y^{2}},g_{xy\left(x^{2}-y^{2}\right)}\right)$ Cooper pairs.
From Eq.~\eqref{eq_strain_components} it
follows that $T_c$ as a function of $\sigma_{\rm 110}$, our experimental control parameter, obeys
\begin{equation}
\Delta T_c\left(\sigma_{{\rm 110}}\right)=\left(\sum_{i=\text{d}, 3}\frac{dT_c}{d\varepsilon_{i}}\alpha_{i}+{\rm sign}\left(\sigma_{{\rm
110}}\right)\left|\frac{dT_c}{d\varepsilon_{6}}\right| \alpha_{6}\right)\sigma_{{\rm 110}}.
\label{eq_Tc_cusp}
\end{equation}
This behavior is illustrated in Fig.~\ref{fig: Tschematic}. For a single component superconducting order parameter and for two-component states other than the ones listed
above, $|dT_c/d\varepsilon_6| = 0$. It is this distinct behavior w.r.t.\ $\varepsilon_{6}$ that allows for the key conclusions  of this paper.

In some cases one expects a second transition at a temperature $T_2 < T_c$ where a composite of the two order parameter components breaks an additional symmetry.
If this happens, one expects behavior similar to Eq.~\eqref{eq_Tc_cusp}, but with the crucial distinction that the sign in front of the cusp is negative: 
\begin{equation}
\Delta T_{2}\left(\sigma_{{\rm 110}}\right)=\left(\sum_{i=\text{d},3}\frac{dT_{2}}{d\varepsilon_{i}}\alpha_{i}-{\rm sign}\left(\sigma_{{\rm
110}}\right)\left|\frac{dT_{2}}{d\varepsilon_{6}}\right| \alpha_{6}\right)\sigma_{{\rm 110}},
\label{eq_T2_cusp}
\end{equation}
as sketched in Fig.~\ref{fig: Tschematic}. From the slopes $|dT_c/d\varepsilon_6|$ and (if a second transition occurs) $|dT_2/d\varepsilon_6|$, an upper bound on the
jump in elastic constant $c_{66}$ may be obtained; this relation is presented below.

\section{Results: measurement of \boldmath{$T_\text{\lowercase{c}}(\sigma_{110})$}}

In order to probe the dependence of $T_c$ on $\sigma_{110}$, we studied the magnetic susceptibility of single crystals of \SRO{}. Stress was applied using piezoelectric-based apparatus that incorporated both force and displacement sensors~\cite{Barber19_RSI}.  Samples were sculpted into dumbell shapes using a Xe plasma focused ion
beam, a step that allows higher stresses to be reached~\cite{Jerzembeck22_NatComm}.  
To measure magnetic susceptibility, concentric coils of a few turns each were wound around the central neck portion of the samples and their mutual inductance was measured.  
For each sample, the zero-stress point was identified by deliberately breaking the sample under tension, then measuring $T_c$ with the two parts separated~\footnote{Susceptibility under $ [110]$ pressure was previously studied \cite{Hicks14_Science} but strain inhomogeneity combined with relatively large pickup coils and a lack of precise knowledge of the zero strain point meant that this could not be regarded as a precision search for a cusp in $T_c$ at zero pressure. 
The steps taken in the current experiment were designed to provide the precision experiment required to either observe the cusp
or place tight upper limits on its magnitude.}.

Three samples were measured. Samples 1 and 2 were taken from the same original crystal, in which the growth direction was almost exactly along $[110]$, while sample 3 was taken
from a crystal where the growth direction was about $15^\circ$ away from $[110]$.  In all cases the samples were cut from the original crystal such that the pressure was applied
along [110] within $\lesssim 3^\circ$.  Samples 1 and 2 both withstood tensile stresses of up to $\sigma_{110} \approx +0.2$~GPa, while sample 3 broke
under very low tensile stress. Samples 1 and 3 were compressed to $\sigma_{110} < -2$~GPa.

For sample 1, there was some hysteresis in $T_c(\sigma_{110})$.  The most likely origin was a hysteretic component of the applied stress that had $\langle 100 \rangle$
principal axes: $T_c$ of \SRO{} responds much more sensitively to $\langle 100 \rangle$ than $\langle 110 \rangle$ shear stress~\cite{Hicks14_Science}. After measurement of
sample~1, the apparatus was modified to attenuate transmission of stress components other than the desired $[110]$ uniaxial stress. This step drastically reduced the
hysteresis for samples 2 and 3. The modification is described in Appendix~\ref{sec_exp_appendix}.

\begin{figure}[ptb]
\includegraphics[width=85mm]{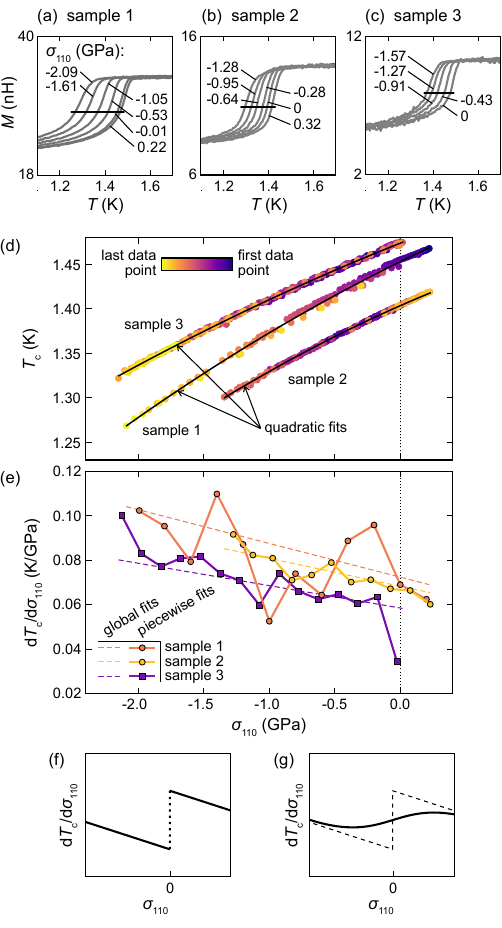}
\caption{(a--c) Temperature dependence of the mutual inductance $M$ of the susceptibility coils wrapped around the sample, for samples 1, 2, and 3, respectively. The numbers
indicate the applied stress, $\sigma_{110}$, in GPa, and $\sigma_{110} < 0$ denotes compression. 
(d) $T_c$, determined as the points where $M$ crossed the thresholds indicated in panels (a--c), against
stress. To illustrate the level of drift, the points are colored by the order in which they were measured. For sample~1, due to hysteresis only data from decreasing-$\sigma$
ramps are shown. The black lines are quadratic fits to the data. (e) Points: $dT_c/d\sigma_{110}$ determined from piecewise linear fits. Straight lines: $dT_c/d\sigma_{110}$ from the quadratic fits in panel d. (f) Expected form of $dT_c/d\sigma_{110}$, if there is a sharp
cusp on top of a quadratic background. (g) Expected form of $dT_c/d\sigma_{110}$ if the cusp is broadened by, for example, internal strain inhomogeneity.}
\label{fig_ACSData}
\end{figure}

Raw data --- the mutual inductance $M$ of the susceptibility coils --- for all three samples are shown in Fig.~\ref{fig_ACSData}(a--c).  For all, the transition width was about
50~mK, and did not increase much as stress was applied.  This transition width could be a consequence of an inhomogeneous defect density, and/or an internal field of $\langle 100
\rangle$ shear strain due to defects.  It is not a consequence of whatever internal field of $\langle 110 \rangle$ shear strain may be present.  It can be seen in
Fig.~\ref{fig_ACSData}(d) that $\sigma_{110} \approx -1$~GPa is required to suppress $T_c$ by 50~mK, which is an unrealistically large internal stress except in the
immediate vicinity of defects~\cite{Ying13_NatComm}.

Fig.~\ref{fig_ACSData}(d) shows $T_c(\sigma_{110})$ of the three samples, where $T_c$ is taken as the point where $M$ crosses the thresholds indicated in panels
(a--c).  To convey the level of drift, the data points are colored by the order in which they were measured.  Due to the hysteresis, for sample~1 only data points taken under
decreasing $\sigma_{110}$ are shown.

Linear fits over the range $-0.2 < \sigma_{110} < 0.2$~GPa give $dT_c/d\sigma_{110} = 68.3$, $65.8$, and $58.5$~mK/GPa for samples 1, 2, and 3, respectively. 
Including uncertainty on the force sensor calibration, we take $\left. dT_c/d\sigma_{110}\right|_{\sigma_{110}=0} = 64 \pm 7$~mK/GPa.

For none of the samples is there an obvious cusp at $\sigma_{110} = 0$.
To look more closely, we determine the slopes $dT_c/d\sigma_{110}$ from piecewise linear fits, choosing windows small enough that for samples 1 and 2 there are windows at $\sigma_{110} > 0$.
Results from this fitting are shown in Fig.~\ref{fig_ACSData}(e).
If there were a sharp cusp in $T_c(\sigma_{110})$ at $\sigma_{110} = 0$, the points at $\sigma_{110} > 0$ would be above the trend from $\sigma_{110} < 0$.
They are not. Also shown in Fig.~\ref{fig_ACSData}(e) is $dT_c/d\sigma_{110}$ determined from quadratic fits to $T_c(\sigma_{110})$ over the entire measured stress range.
The rms difference between the piecewise-fitted and these globally-fitted slopes is 10~mK/GPa.

A cusp could be broadened by internal strain.  In Fig.~\ref{fig_ACSData}(f) we illustrate the expected form of $dT_c/d\sigma_{110}$ if there is a sharp cusp, and in
Fig.~\ref{fig_ACSData}(g) if it is rounded.  With rounding, $dT_c/d\sigma_{110}$ deviates from the background over a range of stress around $\sigma_{110}$, such that even in
sample 3, where essentially no tensile stress could be applied, the effects of a cusp could have become visible.  No such deviation is visible in the data. 

We take a conservative upper limit on any change in slope $dT_c/d\sigma_{110}$ across a cusp, $\Delta(dT_c/d\sigma_{110})$, of 20~mK/GPa, whether the cusp is sharp or rounded. This upper limit
implies 
\begin{equation}
\left| \frac{dT_c}{d\varepsilon_6} \right| = \frac{1}{2\alpha_6} \Delta \left(\frac{dT_c}{d\sigma_{110}}\right) < 1.3\;\text{K}.
\end{equation}
which is our first key experimental result. The implication of this tight upper bound on $|dT_c/d\varepsilon_6|$ will be discussed in
detail later in the paper.

\begin{figure}[ptb]
\includegraphics[width=85mm]{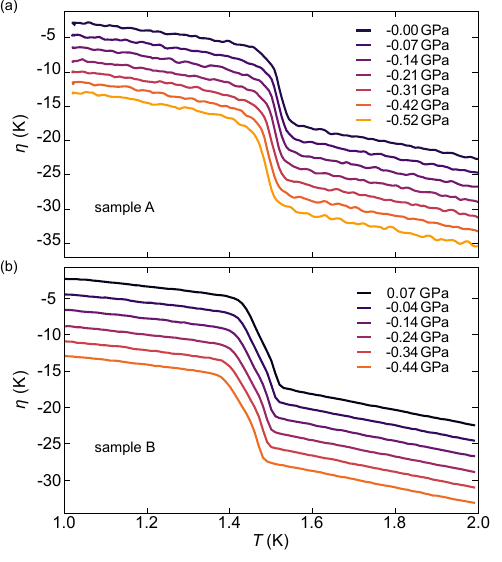}
\caption{\label{fig_ECEData} Elastocaloric coefficient $\eta$ versus temperature for two samples. For clarity, all curves apart from the black ones are shifted vertically with respect to each other. The method of analysis is discussed in the text and Appendix~\ref{Calib_ECE}.}
\label{Fig: elastocaloric}
\end{figure}

\section{Elastocaloric Effect experiments}
\label{ECE}

It is clear from the $T_c(\sigma_{110})$ data that the superconductivity couples much more weakly to shear strain with $\langle 110 \rangle$ than $\langle 100 \rangle$ principal axes; data under $\langle 100 \rangle$ uniaxial stress are published in, for example, Refs.~\cite{Hicks14_Science, Li21_PNAS}. 
It might not then seem that the elastocaloric effect (the change in sample temperature induced by applied stress) under $[110]$ stress would be a particularly effective probe of the physics. However, that is not necessarily the case. 
Under adiabatic conditions, the elastocaloric coefficient $\eta$ is given by 
\begin{equation}
\eta \equiv \frac{d T}{d \varepsilon} = -\frac{T}{C} \frac{\partial S}{\partial \varepsilon},
\label{Eq:elastocaloric}
\end{equation}
where $C$ is the heat capacity, $S$ the entropy, and the direction of strain $\varepsilon$ is fixed by the conditions of the experiment. Suppose, for a moment, single-component superconductivity. 
The Ginzburg-Landau free energy of the single-component superconducting state is then given by
\begin{equation}
F = F_n + \frac{a(T, \varepsilon)}{2}|\psi|^2 + \frac{u}{4}|\psi|^4.
\end{equation}
Here, $a(T, \varepsilon) = a_0(T - T_\text{c0}(\varepsilon))$, and $F_n$ is the free energy in the normal state. 
The entropy below $T_\text{c0}$ is given by
\begin{equation}
S(T, \varepsilon) = S_n + \frac{a_0(T - T_\text{c0}(\varepsilon))}{2u},
\end{equation}
where $S_n = -\partial F_n/\partial T$ is the normal state entropy. 
It follows from Eq.~\ref{Eq:elastocaloric} that the change in $\eta$ at $T_c$, $\Delta \eta$, equals
\begin{equation}
\frac{\Delta\eta}{\eta_n(T_c)} = -\left(1 - \frac{1}{\eta_n(T_c)}\frac{dT_c}{d\varepsilon}\right) \times \left(1 + \frac{C_n(T_c)}{\Delta C}\right)^{-1},
\label{eq_jump_ECE}
\end{equation}
where $\eta_n$ is the normal-state elastocaloric coefficient, $C_n$ is the normal-state heat capacity, and $\Delta C$ is the heat capacity jump at $T_c$. 
In the case of two-component order and near zero stress, Eq.~\ref{eq_jump_ECE} applies for strains $\varepsilon$ that transform under the trivial representation: $\varepsilon_d$ or $\varepsilon_3$. 
For two-component order and under non-zero stress that splits the transitions, Eq.~\ref{eq_jump_ECE} applies for strains $\varepsilon_d$, $\varepsilon_3$, and $\varepsilon_6$.

Eq.~\eqref{eq_jump_ECE} allows for rather general insights
into the strain dependence of $T_c$.  Suppose the term $dT_c/d\varepsilon$ is negligible compared to $\eta_n$.  Since $1+C_n/\Delta C>1$, the magnitude
of $\eta$ would fall at $T_c$, but it would not change sign. This behavior is in contrast to the behavior under $[100]$ uniaxial stress, where $dT_c/d\varepsilon$ is
much larger and $\eta$ is observed to change sign at $T_c$~\cite{Li22_Nature}.  Importantly, Eq.~\ref{eq_jump_ECE} shows that $\Delta \eta$ can be substantial even if
$dT_c/d\varepsilon = 0$. 

It has been demonstrated that for \SRO{} under uniaxial stress at low temperatures, $\eta$ can be measured with a higher signal-to-noise ratio than heat capacity~\cite{Li22_Nature,
Li21_PNAS}.  Therefore, in the case that the superconducting transition splits into transitions at $T_c$ and $T_2$, the elastocaloric effect is an ideal probe to search for
thermodynamic signatures of the transition at $T_2$. 

Elastocaloric effect data from two samples under $[110]$ uniaxial pressure are shown in Fig.~\ref{Fig: elastocaloric}. In the normal state, $\eta$ is negative over the range
of pressures and temperatures studied, meaning that $S$ in the normal state increases when samples are tensioned. Inspection of the data shows that, at
all strains measured, only one transition is observed. There is no visible sign of uniaxial pressure-dependent splitting of the main transition--- any structure in the transition
that is seen at zero pressure (likely due to slight inhomogeneity of the strain field and/or defect density) remains the same at nonzero pressure. We may, therefore, proceed
with analysis under a tentative hypothesis of single-component order. The observed behavior across $T_c$ is as expected for small $|dT_c/d\varepsilon_{110}|$: $\eta$ is smaller
below $T_c$, but its sign is unchanged.

The small size of the signal makes data analysis more challenging than for $[100]$ uniaxial stress~\cite{Li22_Nature}.  To analyze the data, we assume that the normal-state heat
capacity of \SRO{} at low temperatures is given by:
\begin{equation}
C_n = (\gamma_0 + \gamma_1 \varepsilon_{110})T + \beta T^3,
\label{eq_normalStateHeatCapacity}
\end{equation}
which yields
\begin{equation}
\eta_n = \frac{-\gamma_1 T}{\gamma_0 + \gamma_1 \varepsilon_{110} + \beta T^2}.
\label{eq_normalStateECE}
\end{equation}
The elastocaloric effect is a recently-introduced technique~\cite{Ikeda19_RSI}, and experimental uncertainties remain, especially in quantifying the actual strain oscillation
amplitude, $\delta \varepsilon$, and the degree of adiabaticity, $A$. ($A=1$ denotes perfect adiabaticity, and $A=0$ denotes complete dissipation of temperature oscillations into
the stress cell.) However, the key quantity in Eq.~\ref{eq_jump_ECE}, $\Delta \eta / \eta_n(T_c)$, can be obtained directly from the jump in measured thermocouple voltage at $T_c$,
without knowledge of $A$ or $\delta \varepsilon$. That quantity can then be used to solve for $\gamma_1$, which is the only quantity in Eqs.~\ref{eq_normalStateHeatCapacity}
and~\ref{eq_normalStateECE} that is not fixed by experimental data. We set $\gamma_0$, $\beta$, and $\Delta C/C_n$ to literature values of 37.5~mJ/mol-K$^2$, 0.197~mJ/mol-K$^4$,
and 0.65, respectively~\cite{Nishizaki99_JLTP, Ghosh21_NatPhys}. In Eq.~\ref{eq_jump_ECE}, we set $\varepsilon$ to be $\varepsilon_{110}$, corresponding to our measurement
conditions, and set $dT_c/d\varepsilon$ to $64\;\text{mK/GPa} \times 187\;\text{GPa} = 12.0\;\text{K}$. Averaging results from the two samples, we obtain $\gamma_1 =
0.43$~J/mol-K$^2$.

All the terms in Eq.~\ref{eq_normalStateECE} are now known, and we may then apply Eq.~\ref{eq_normalStateECE} to extract $A(T) \times \delta \varepsilon(T)$. 
In effect, where the raw data deviate from the form given in Eq.~\ref{eq_normalStateECE}, we make an assumption that it is more likely to be due to $T$ dependence of $A \delta \varepsilon$ than that there are terms in $C_n$ beyond those in Eq.~\ref{eq_normalStateHeatCapacity} that are important at low $T$ and $\varepsilon_{110}$. 
We then extrapolate $A(T) \delta \varepsilon(T)$ to $T < T_c$ to obtain a best estimate for the ECE in the superconducting state. 
This is what is shown in Fig.~\ref{Fig: elastocaloric}. Further details are given in Appendix~\ref{Calib_ECE}.

\section{Discussion}

The qualitative finding from these $\langle 110 \rangle$ uniaxial pressure experiments on \SRO{} is that we do not observe either of two key predicted features of
two-component superconducting states: neither a cusp in $T_c$, nor splitting of the superconducting transition into two transitions at $T_c$ and $T_2$.  In this
discussion, we review the consistency of these findings with those of other experiments, and the implications for theories of the superconducting order parameter of \SRO. In both
sections, we frame the discussion with the quantitative bounds that we have placed on the putative existence of a cusp or of transition splitting, rather than on categorical
statements that neither exist.

\subsection{Comparison with hydrostatic pressure dependence of $T_c$}

In this work we have determined that $dT_c/d\sigma_{110} = 64\pm 7$~mK/GPa.  As shown in Appendix~\ref{Comp_appendix}, combining this result with previous measurement under 
pressure applied along the $c$-axis that $dT_c/d\sigma_{001} = 76\pm 5$~mK/GPa \cite{Jerzembeck22_NatComm} enables a prediction for the dependence of $T_c$ on
hydrostatic pressure $\sigma_\text{hyd}$: $dT_c/d\sigma_\text{hyd} = 204\pm 12$~mK/GPa.  This allows a useful cross-check on the accuracy of our uniaxial pressure data,
because the dependence of $T_c$ on hydrostatic pressure has been measured in four independent experiments \cite{Forsythe02_PRL, Shirakawa97_PRB, Svitelskiy08_PRB, Grinenko21_NatComm}. 
However, since the $T_c$ of most studied samples were substantially below 1.5~K, which is pointing to strong disorder, we focus on the hydrostatic pressure dependence of Ref.~\cite{Forsythe02_PRL}, optimal-$T_c$ experiment, which found $\sigma_\text{hyd}$: $dT_c/\sigma_\text{hyd} = 220 \pm 20$~mK/GPa.
The agreement with the derivation from uniaxial stress measurements is reassuring.

\subsection{Comparison with $\mu$SR under \boldmath{$[110]$} uniaxial pressure}

Recently, a $\mu$SR study under $\langle 110 \rangle$ uniaxial pressure inferred transition splitting, with a TRSB transition at $T_2 = T_c - (0.7 \pm 0.2)$~K/GPa
\cite{Grinenko23_PRB}, which corresponds to $dT_{2}/ d\varepsilon_{{\rm 110}} \approx131\,{\rm K}$. 
The transition at $T_2$ would be within our measured temperature range for each of the stresses shown for sample B in Fig.~\ref{Fig: elastocaloric}, and
would have been visible as long as $|\Delta \eta_2|$ were larger than $\sim 0.2$~K, or 2\% of the jump in ECE at $T_c$, $|\Delta \eta_c|$. As with previous
comparisons of $\mu$SR and heat capacity data under $[100]$ uniaxial pressure~\cite{Grinenko21_NatPhys}, it is hard to imagine a transition to a second homogeneous thermodynamic
state yielding such a small anomaly.
We note the low statistical significance of the splitting reported in Ref.~\cite{Grinenko23_PRB}; a simple repeat of that measurement would be useful. 
However, we believe that the direction of travel is toward fundamental re-evaluation of the interpretation of $\mu$SR data in unconventional superconductors.

\subsection{Relationship with jumps in $c_{66}$ observed in ultrasound experiments}

\begin{figure}[ptb]
\includegraphics[width=85mm]{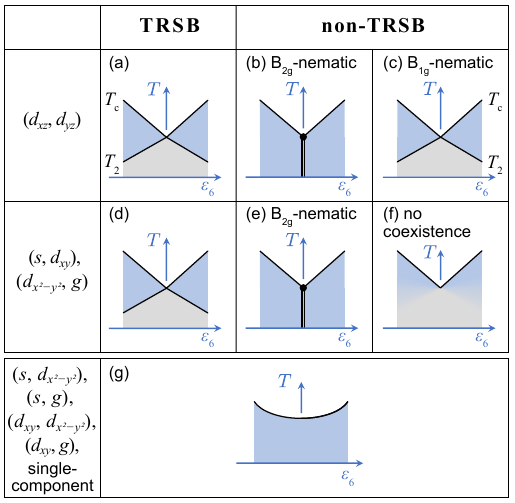}
\caption{$T$-$\varepsilon_6$ phase diagrams for various possible order parameters. $g$ denotes $g_{xy(x^2-y^2)}$. The first row illustrates $(d_{xz}, d_{yz})$ order and the
ways these components can combine at temperatures well below $T_c$. In panel (a), they form TRSB order, $d_{xz} \pm id_{yz}$. In panel (b), B\textsubscript{2g}-nematic order,
$d_{xz} \pm d_{yz}$. In panel (c), they form B\textsubscript{1g}-nematic order: condensation of either component alone, without coexistence.  The next row illustrates the equivalent possibilities for accidentally degenerate two-component orders that couple linearly to $\varepsilon_6$. The bottom row illustrated the expected strain dependence of order parameters, which do not couple linearly to $\varepsilon_6$. In all panels, single black lines indicate second-order transitions, double lines indicate first-order transitions, and color gradients indicate crossovers. Further explanation is provided in the text.}
\label{fig_schematicPhaseDiagrams}
\end{figure}

One of the motivations for the current experiments was to perform a careful comparison with the results of ultrasound experiments, which have resolved jumps in the elastic constant
$c_{66}$ at the superconducting transition.  The observation of a jump in this elastic constant is particularly significant because it implies the existence of some kind
of two-component superconducting order parameter; a single-component order parameter gives a jump in other elastic constants, but not in $c_{66}$.  Jumps of 0.03~MPa and 0.15~MPa
were reported on the basis of separate pulse-echo measurements at 169 MHz and 201 MHz respectively \cite{Benhabib21_NatPhys}, and of 1.05 MPa on the basis of a resonant ultrasound
experiment performed at much lower frequencies of approximately 2 MHz \cite{Ghosh21_NatPhys}.  The difference between the two pulse-echo results was attributed to possible mode
mixing in the 201 MHz experiment.  It has been suggested that the difference between the pulse-echo and resonant ultrasound results is a consequence of the very different
measurement frequencies, with the higher frequencies thought to suppress the jump from its intrinsic thermodynamic value \cite{Benhabib21_NatPhys}.  For the purposes of the
analysis that follows, we will take the quoted numbers at face value and examine the extent to which they are consistent with our results, under an assumption that all the
experiments are giving information on bulk, homogeneous thermodynamic phases.  

To frame our thermodynamic analysis, we first give some important results about Ehrenfest relations, giving the full derivation in Appendix~\ref{sec_GL_appendix}. Broadly speaking,
there are two possibilities for two-component order parameters that give a jump in $c_{66}$ in \SRO{}. One is that degeneracy of the components is symmetry-protected, i.e., the
components are equivalent on a tetragonal lattice. Given the strong evidence for even-parity, spin-singlet superconductivity in \SRO~\cite{Palle23_PRB, Pustogow19_Nature,
Chronister21_PNAS, Li22_Nature}, these two components would have $d_{xz}$ and $d_{yz}$ symmetry.  The other possibility is accidental degeneracy between certain
non-symmetry-related components, namely the pair $(s, d_{xy})$ or the pair $(d_{x^2-y^2}, g_{xy(x^2-y^2)})$. For both pairs, multiplying the components yields a composite order
that couples linearly to shear strain with $\langle 110 \rangle$ principal axes, that is, $\varepsilon_6$.  Single-component order parameters in contrast do not yield a jump in
$c_{66}$ at $T_c$.  Neither do two-component order parameters which do not couple to $\varepsilon_6$ shear strain:  the pairs $(s, d_{x^2-y^2})$, $(s,
g_{xy(x^2-y^2)})$, $(d_{xy}, d_{x^2-y^2})$, and $(d_{xy}, g_{xy(x^2-y^2)})$.  A table of these possibilities is shown in Fig.~\ref{fig_schematicPhaseDiagrams}.

Consider first the symmetry-protected possibility, $(d_{xz}, d_{yz})$. 
If, at temperatures well below $T_c$, the components combine to break time reversal symmetry, then under shear strain with $\langle 110 \rangle$ principal axes the transition would split into a transition at $T_c$ into $d_{xz} \pm d_{yz}$ order, followed by a transition at $T_2$ into $(d_{xz}
\pm d_{yz}) \pm i(d_{xz} \mp d_{yz})$ order. This possibility is illustrated in Fig.~\ref{fig_schematicPhaseDiagrams}(a). If the components combine to form
B\textsubscript{2g}-nematic order (that is, $d_{xz} \pm d_{yz}$ order), there would be no second transition at lower temperature. There would be a first-order transition along the
$\varepsilon_6 = 0$ line, where the favored orientation of the nematicity flips. This possibility is illustrated in Fig.~\ref{fig_schematicPhaseDiagrams}(b).  Finally, if the
components form B\textsubscript{1g}-nematic order, i.e.\ $d_{xz}$ or $d_{yz}$ without coexistence of the two components, then $\langle 110 \rangle$ shear strain is again expected
to yield a split transition. The order parameter just below $T_c$ would be $d_{xz} \pm d_{yz}$, and at $T_2$ there would be a symmetry-breaking transition at which the
principal axes of the nematicity rotate away from the $\langle 110 \rangle$ axes and towards the $\langle 100 \rangle$ axes.

The accidentally-degenerate pairs $(s, d_{xy})$ and $(d_{x^2-y^2}, g_{xy(x^2-y^2)})$ could also combine to yield TRSB or B\textsubscript{2g}-nematic orders -- for example, $s \pm
id_{xy}$ or $s \pm d_{xy}$. 
The $T$-$\varepsilon_6$ phase diagrams would be qualitatively the same as for the $(d_{xz}, d_{yz})$ pair; see
Figs.~\ref{fig_schematicPhaseDiagrams}(d, e). 
The remaining possibility --- absence of co-existence --- is qualitatively different from the $(d_{xz}, d_{yz})$ case, 
because for these accidentally-degenerate pairs there is no diagonal reflection symmetry $x \leftrightarrow y$ which protects the B\textsubscript{2g} nematic state.
Just below $T_c$ and with $\varepsilon_6 \neq 0$, the two components would combine to yield B\textsubscript{2g}-nematic order, and a cusped dependence of $T_c$ on $\varepsilon_6$. 
As $T$ is further reduced, one of the components would come to dominate, but this would be a
smooth crossover. This possibility is illustrated in Fig.~\ref{fig_schematicPhaseDiagrams}(f).

Analysis of each situation yields the following Ehrenfest relations [Appendix~\ref{sec_Ehrenfest_relations_appendix}].
For B\textsubscript{2g}-nematic order constructed from $(d_{xz}, d_{yz})$ components,
\begin{equation}
\Delta c_{66} = \frac{\Delta C_0}{T_\text{c0}} \bigg | \frac{dT_c}{d\varepsilon_6} \bigg |^2,
\label{eq_OneComponentEhrenfest}
\end{equation}
where $\Delta C_0$ is the jump in heat capacity at the superconducting transition in the unstressed system, and $T_\text{c0}$ is $T_c$ in the unstressed system. On the other
hand, for both $d_{xz} \pm id_{yz}$ TRSB and $(d_{xz}, d_{yz})$ B\textsubscript{1g}-nematic order,
\begin{equation}
\Delta c_{66} =  \frac{\Delta C_0}{T_\text{c0}} \left|\frac{dT_c}{d\varepsilon_6}\right| \left| \frac{dT_2}{d\varepsilon_6} \right|.
\label{eq_DegTwoComponentEhrenfest}
\end{equation}
This equation is obtained from Eq.~\eqref{eq_fullEhrenfestRelation} in Appendix~\ref{sec_GL_appendix}, with $\alpha$ in that equation set to zero.
Here, we label as $\Delta C_0$ the heat capacity jump without splitting--- anywhere along the transition line when splitting does not occur, and at $\varepsilon_6=0$ when it does. $T_\text{c0}$ is the critical temperature in the absence of splitting.

We begin with the simpler of the above Ehrenfest relations, Eq.~\eqref{eq_OneComponentEhrenfest}, which applies to $d_{xz} \pm d_{yz}$ B\textsubscript{2g}-nematic order. $\Delta C_0$ in \SRO{} is 38 mJ/mol-K \cite{Nishizaki00_JPSJ, Deguchi04_JPSJ}.
Combining this with the largest and smallest literature values for $\Delta c_{66}$, 1.05 MPa and 0.03 MPa, yields  values for
$|dT_c/d\varepsilon_6|$ of 48~K and 8~K, respectively.  
As stated above, we resolve no cusp, with our experiment placing an upper limit on $|dT_c/d\varepsilon_6|$ of 1.3 K.  
If both our experiments and the ultrasound studies are probing bulk, homogeneous thermodynamic states, there is therefore a discrepancy of between a factor of 37 and 6 between our upper limit on $|dT_c/d\varepsilon_6|$ and predictions from Eq.~\eqref{eq_OneComponentEhrenfest} using measured values of $\Delta c_{66}$. 
We can therefore rule out bulk B\textsubscript{2g}-nematic order, namely $d_{xz} \pm d_{yz}$ order, as the origin of the observed jumps in $c_{66}$.

In the case of $d_{xz} \pm id_{yz}$ or B\textsubscript{1g}-nematic $d_{xz}$ or $d_{yz}$ order, where it is Eq.~\eqref{eq_DegTwoComponentEhrenfest} that applies, the very small
upper limit that our measurements place on $|dT_c/d\varepsilon_6|$ could in principle be compensated by a very large value of $|dT_2/d\varepsilon_6|$. However, the numbers
are stark. For $\Delta c_{66} = 0.03$~MPa,  $|dT_\text{2}/d\varepsilon_6| > 51$~K would be required to be compatible with our finding that $|dT_c/d\varepsilon_6| < 1.3$~K.
For the largest measured value of $\Delta c_{66}$, 1.05~MPa, $|dT_\text{2}/d\varepsilon_6| > 1785$~K would be required.

\begin{figure}[t]
\includegraphics[width=\columnwidth]{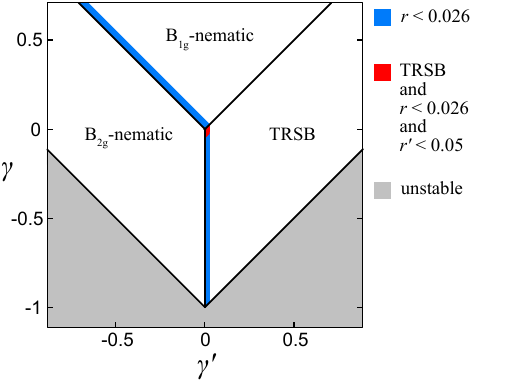}
\caption{Phase diagram of the Ginzburg-Landau parameter space for $(d_{xz}, d_{yz})$ superconductivity, based on the free energy written in Eq.~\eqref{eq_symmProtectedGL}.
$\gamma$ and $\gamma^\prime$ are parameters of this free energy that define whether the ground state is B\textsubscript{1g}-nematic, B\textsubscript{2g}-nematic, or TRSB.
The blue region is the region consistent with the condition $|dT_c/d\varepsilon_6| < 1.3$~K and $\Delta c_{66} > 0.03$~MPa.
Within the TRSB region, the red patch near the  $\gamma = \gamma' = 0$ triple point is additionally consistent with the heat capacity data of Ref.~\cite{Li21_PNAS} under $\langle 100\rangle$ uniaxial stress.
 The upper bounds on the ratios $r$, $r^\prime$ are explained in the main text.}
\label{fig_triplePointTuning}
\end{figure}

A difference of this magnitude between $|dT_\text{2}/d\varepsilon_6|$ and $|dT_c/d\varepsilon_6|$ would require a very high degree of tuning.
To quantify this fine-tuning, let us parameterize the free energy of $(d_{xz}, d_{yz})$ order in the following way [Eq.~\eqref{eq_F_Delta}]:
\begin{equation}
\begin{gathered}
F = \frac{a}{2} \big(|\Delta_1|^2 + |\Delta_2|^2\big) + \frac{u}{4} \big(|\Delta_1|^2 + |\Delta_2|^2\big)^2 \\
+ \gamma u |\Delta_1|^2 |\Delta_2|^2 + \gamma' \frac{u}{2} \big(\Delta_1^{*} \Delta_1^{*} \Delta_2 \Delta_2 + \Delta_2^{*} \Delta_2^{*} \Delta_1 \Delta_1\big).
\end{gathered} \label{eq_symmProtectedGL}
\end{equation}
$\Delta_1$ and $\Delta_2$ are, respectively, the amplitudes of the $d_{xz}$ and $d_{yz}$ components. $\gamma^\prime > 0$ favors combining them with a $\pi/2$ phase shift, yielding
$d_{xz} \pm id_{yz}$ TRSB order, while $\gamma^\prime < 0$ favors $d_{xz} \pm d_{yz}$ B\textsubscript{2g}-nematic order.
$\gamma > 0$ disfavors coexistence of the two components; B\textsubscript{1g}-nematic order is obtained when $\gamma > |\gamma^\prime|$.
For a generic microscopic theory, the three quartic coefficients are expected to be comparable in magnitude, giving $\gamma$, $\gamma^\prime$ on the order of $1$.
The $\gamma$-$\gamma^\prime$ phase diagram is shown in Fig.~\ref{fig_triplePointTuning}.

As discussed above, our data are not consistent with B\textsubscript{2g}-nematic order, but TRSB and B\textsubscript{1g}-nematic orders are in principle possible.
By exploiting the Ehrenfest relation~\eqref{eq_DegTwoComponentEhrenfest}, one may bound the ratio
\begin{equation}
r \equiv \left|\frac{dT_c}{d\varepsilon_6}\right| \left|\frac{dT_2}{d\varepsilon_6}\right|^{-1} \label{eq_r}
\end{equation}
from above to obtain $r < (1.3\;\text{K})/(51\;\text{K}) = 0.026$. 
Here we have used the smallest reported $\Delta c_{66} > 0.03$~MPa, to obtain a conservative estimate.
One may show that in the B\textsubscript{1g}-nematic region $r = \gamma + \gamma'$, and that in the TRSB region $r = 2 \gamma' / (1 + \gamma - \gamma')$ [Eq.~\eqref{eq_ratioRelationsAppendix}].
Therefore, the regions of the phase diagram in Fig.~\ref{fig_triplePointTuning} colored blue would be consistent with our data: the interaction between the two components must be tuned right to the cusp of B\textsubscript{2g} nematicity, but without the order being B\textsubscript{2g}-nematic.

We can add to this diagram constraints imposed by measurements under $\langle 100 \rangle$ uniaxial stress.
The analogous ratio
\begin{equation}
r^\prime \equiv \left|\frac{dT_c}{d(\varepsilon_1 - \varepsilon_2)}\right| \left|\frac{dT_2}{d(\varepsilon_1 - \varepsilon_2)}\right|^{-1} \label{eq_rprime}
\end{equation}
is inversely related to the ratio of the heat capacity jumps at the upper and lower transitions [Eq.~\eqref{eq_ratioRelationsAppendixB1g}].
In Ref.~\cite{Li21_PNAS}, it was shown that the heat capacity anomaly at any second transition is at most 5\% of that at $T_c$, corresponding (again, for $d_{xz} \pm id_{yz}$ order) to the condition  $r^\prime < 0.05$.
Within the TRSB region of the phase diagram, $r' = - (\gamma-\gamma') / (1+\gamma-\gamma')$.
Therefore, the region within the TRSB phase consistent with both sets of experiments is only the red region in Fig.~\ref{fig_triplePointTuning}.
That is, if the order parameter of \SRO{} is a TRSB order parameter constructed from components with symmetry-protected degeneracy, the interaction between these components would need to be doubly fine-tuned so as to lie extremely close to the triple point in the phase diagram.

So far, there are no data ruling out B\textsubscript{1g} nematicity at the same level that B\textsubscript{2g} nematicity is excluded here. The upper limit on
$|dT_c/d(\varepsilon_1-\varepsilon_2)|$ is 11~K~\cite{Watson18_PRB}, not as tight as the limit set here on $|dT_c/d\varepsilon_6|$.
Therefore, within the B\textsubscript{1g}-nematic region we require only the single level of fine-tuning described above, $\gamma + \gamma^\prime < 0.026$, but emphasize that it is a stringent condition.

In case of TRSB order constructed from accidentally-degenerate components, the Ehrenfest relation Eq.~\eqref{eq_DegTwoComponentEhrenfest} becomes an inequality:
\begin{equation}
\Delta c_{66} \leq \frac{\Delta C_0}{T_\text{c0}} \left|\frac{dT_c}{d\varepsilon_6}\right| \left| \frac{dT_2}{d\varepsilon_6} \right|,
\label{eq_AccTwoComponentEhrenfest}
\end{equation}
and we still obtain an upper bound on $\Delta c_{66}$ derivable from the strain variations of the transition temperatures. This equation is Eq.~\eqref{eq_fullEhrenfestRelation}
from Appendix~\ref{sec_Ehrenfest_relations_appendix}, with $\alpha \neq 0$. As under the $d_{xz} \pm id_{yz}$ hypothesis, reconciling our data with the observed $\Delta c_{66}$ would still require fine tuning, with
$|dT_2/d\varepsilon_6|$ vastly larger than $|dT_c/d\varepsilon_6|$. 

In the absence of co-existence --- the case illustrated in Fig.~\ref{fig_schematicPhaseDiagrams}(f) --- no similarly strong conclusions can be drawn.
Eq.~\eqref{eq_OneComponentEhrenfest} would apply directly at the superconducting transition, as it must, but under the hypothesis of accidental degeneracy there would be a crossover
just below $T_c$ from strain-induced B\textsubscript{2g} nematicity to single-component order. In experimental data, it is likely that the changes in
$c_{66}$ from the two features would merge into a single resolution-limited step, so that Eq.~\eqref{eq_OneComponentEhrenfest} would not be applicable in practice.  Nevertheless, it
seems near-certain that reconciling observed values of $\Delta c_{66}$ with our upper limit on $|dT_c/d\varepsilon_6|$ would still require a high degree of tuning.

\subsection{Implications of the lack of transition splitting in elastocaloric measurements} 

The lower limit $|dT_\text{2}/d\varepsilon_6| > 51$~K established above for $d_{xz} \pm id_{yz}$ and B\textsubscript{1g}-nematic $(d_{xz}, d_{yz})$ order is made even more stringent by analysis of our elastocaloric effect data. For small splitting, i.e.  $T_c - T_\text{2} \ll T_{c0}$, we require
\begin{equation}
\frac{\Delta C_c}{T_c} + \frac{\Delta C_2}{T_2} = \frac{\Delta C_0}{T_{c0}},
\end{equation} 
where $\Delta C_c$ and $\Delta C_2$ are respectively the jumps in heat capacity at $T_c$ and $T_2$. 
The condition that degeneracy of the two components is symmetry-protected imposes the additional condition that 
\begin{equation}
\frac{\Delta C_c}{T_c}\; \frac{T_2}{\Delta C_2} = \left|\frac{dT_2}{d\varepsilon_6}\right| \left|\frac{dT_c}{d\varepsilon_6}\right|^{-1}.
\label{eq_slopeCondition}
\end{equation}

Making use of these conditions, it is straightforward to derive the ECE. The derivation is shown in Appendix~\ref{Jump_ECE}.  In Fig.~\ref{fig_ECEExpectations}, we show the range
of $|dT_c/d\varepsilon_6|$ and $|dT_2/d\varepsilon_6|$ over which a second transition at $T_2$ would have been observable in our elastocaloric effect data.  To obtain this
plot, we set a conservative observability threshold that the jump in $\eta$ at $T_2$ be at least 0.2~K. Although the noise level for sample B is smaller than this, inhomogeneity
broadening might be larger than for the transition at $T_c$, due to its potentially steeper slope.  This observability threshold yields the curved line that bounds the
observable region on the left.  Added to this plot is the relation between $|dT_c/d\varepsilon_6|$ and $|dT_2/d\varepsilon_6|$ fixed by the constraint $\Delta c_{66} = 0.03$~MPa.
It crosses out of the observable region at $|dT_2/d\varepsilon_6| = 144$~K. That is, to account for non-observation of a second transition in the ECE data under an assumption of
$(d_{xz}, d_{yz})$ order and taking $\Delta c_{66} = 0.03$~MPa, we require $|dT_c/d\varepsilon_6| < 0.47$~K and $|dT_2/d\varepsilon_6| > 144$~K, which is a substantial tightening
of the condition $|dT_c/d\varepsilon_6| < 1.3$~K obtained from direct measurement of $T_c(\varepsilon_{110})$.

The upper and lower boundaries of the observable region correspond to a requirement:
\[
42\;\text{K} < \left|\frac{dT_2}{d\varepsilon_{110}}\right| < 750\;\text{K}.
\]
The lower limit corresponds to a change of 0.1~K at the largest strain applied to sample~B, $-0.45\;\text{GPa}/187\;\text{GPa} = -2.4 \cdot 10^{-3}$. 
At this strain, $T_2$ would be $\approx 0.07$~K less than $T_c$, which we estimate as the minimum separation at which a second transition would become distinct from the main transition. 
The upper bound corresponds to a requirement that $T_2$ not drop below our lowest measurement temperature, 1~K, at the smallest applied strain, $\approx 0.10\;\text{GPa} / 187\;\text{GPa} = 5.3 \cdot 10^{-4}$. 
The $|dT_2/d\varepsilon_6|$-$|dT_c/d\varepsilon_6|$ curve that would yield $\Delta c_{66} = 1.05$~MPa lies entirely above the observable region in Fig.~\ref{fig_ECEExpectations}, and therefore, if we take $\Delta c_{66} = 1.05$~MPa, our ECE data do not further tighten the constraint $|dT_c/d\varepsilon_{6}| < 1.3$~K obtained from direct measurement.

This analysis does not strictly apply to fine-tuned accidental degeneracies, but the situation for those is expected to be similar. 
There might be alternative fine-tuning routes that yield small $\Delta C_2$ with modest $|dT_2/d\varepsilon_6|$, but our data nevertheless imply at least two levels of fine-tuning
with accidental degeneracy: the accidental degeneracy itself, and the tuning required to obtain the observed $\Delta c_{66}$ in a way consistent with the null results in this
report.

\begin{figure}[ptb]
\includegraphics[width=85mm]{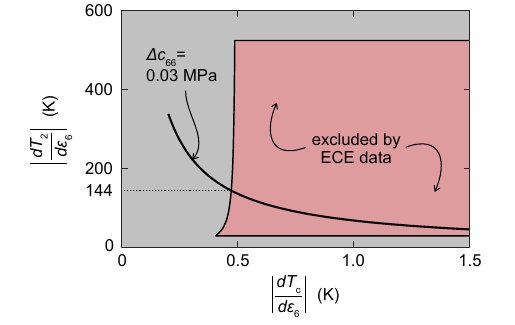}
\caption{\label{fig_ECEExpectations} Region of observability of a second transition in elastocaloric effect data. In the red region, a jump in the elastocaloric effect would have
been resolvable at $T_2$ in our data, under an assumption of $d_{xz} \pm id_{yz}$ or B\textsubscript{1g}-nematic, $(d_{xz}, d_{yz})$ superconductivity.  
The thick line is the set of points defined by the condition $\Delta c_{66} = 0.03$~MPa.} 
\end{figure}

\subsection{Implications for understanding ultrasound data}

The above analysis, combined with the fact that the literature values of $\Delta c_{66}$ have a large variation, causes us to speculate that all the reported ultrasound measurements of $\Delta c_{66}$ substantially exceed the value (which may well be zero) that can be attributed to a homogeneous, two-component order parameter alone.  
We propose as subjects for future study the effects on the ultrasound measurements of quenched disorder, particularly extended defects, in the crystals.  
Possible inhomogeneities of the superconducting state itself should also be carefully considered.
For example, there is evidence for domain walls in the superconducting state~\cite{Kidwingira06_Science, Anwar13_SciRep, Ghosh22_PRB}, and their motion, which was not considered here, might affect
ultrasound measurements. We note in passing that attributing the variation in the ultrasound results to differences in the measurement frequency does not help to resolve the
discrepancies that our work has highlighted.  The reverse is true, because it is more likely that lower-frequency measurements would yield the thermodynamically correct result, but
it is the lowest-frequency measurement that yielded the largest $\Delta c_{66}$.  Finally, we note that the limit  $|dT_c/d\varepsilon_6| < 1.3$~K is independent of anything
concluded from Eqs.~\ref{eq_DegTwoComponentEhrenfest}-\ref{eq_AccTwoComponentEhrenfest}.  Within the standard Ginzburg-Landau theory described in Appendix~\ref{sec_GL_appendix},
this limit is satisfied automatically by any single-component order parameter, for which $|dT_c/d\varepsilon_6| = 0$.  For theories involving two-component order parameters it sets a key constraint, along with the others regarding the observability of transition splitting in ECE measurements.  These constraints are stringent, and it remains to be seen
whether they can be met by any plausible microscopic theory of a two-component order parameter.

\section{Conclusion}
We have measured $T_c$ and the elastocaloric effect in \SRO{} under uniaxial stress applied along the crystalline $[110]$ direction, and found no sign of a
two-component superconducting state.  Namely, we found neither a cusp in the strain dependence of $T_c$ around zero strain, nor a second transition in elastocaloric effect
data under nonzero applied strain.  To reconcile our data with even the smallest reported jump in $c_{66}$ under a hypothesis of homogeneous two-component superconductivity
requires an extreme level of tuning for all proposed order parameters, while some are effectively ruled out.  The difficulty in obtaining clear thermodynamic evidence for
two-component superconductivity, both in the results reported here and in previous measurements under $[100]$ uniaxial stress, means that the possibility that the superconducting
order parameter of \SRO{} is single-component, without breaking time reversal symmetry, must be seriously considered.  The true experimental conditions and/or the
interpretations of the measurement results, both thermodynamic and non-thermodynamic, that have shown evidence for time reversal symmetry breaking should be re-investigated.
However, it is a fact that a large number of experimental probes have found unusual behavior in \SRO{}, such as the jump in $c_{66}$, increased ultrasound dissipation below
$T_c$~\cite{Ghosh22_PRB}, nonzero Kerr rotation~\cite{Xia06_PRL}, and anomalous switching behavior in junctions~\cite{Kidwingira06_Science, Nakamura12_JPSJ}, among others.
Therefore, even if the bulk order parameter turns out to be both spin-singlet and single-component, it appears probable that there is nevertheless something unusual about the
superconductivity of \SRO{}, and that the nature of this "unusualness" has perhaps not been identified even in approximate form by the research community.

\begin{acknowledgments}
G.P. and J.S. were supported by the German Research Foundation (DFG) through CRC TRR 288 "ElastoQMat" project B01. C.W.H. acknowledges support from the Engineering and Physical
Sciences Research Council (U.K.) (EP/X012158/1). J.S. acknowledges support by a Weston Visiting Professorship at the Weizmann Institute of Science, where part of this work was
performed.  T.S. and M.B. acknowledge the support of NSERC, in particular the Discovery Grant [RGPIN-2020-05842], the Accelerator Supplement [RGPAS-2020-00060], and the Discovery
Launch Supplement [DGECR-2020-00222].  N.K. is supported by JSPS KAKENHI (Nos. JP18K04715, JP21H01033, JP22K19093, and 24K01461).  
\end{acknowledgments}

\section*{Data Availability}

The data that support the findings of this study are openly available
from the Max Planck Digital Library~ \footnote{https://doi:10.17617/3.VX9KIR}.

\appendix

\section{Sample carrier design for low hysteresis} \label{sec_exp_appendix} 
\begin{figure}[tb]
\includegraphics[width=85mm]{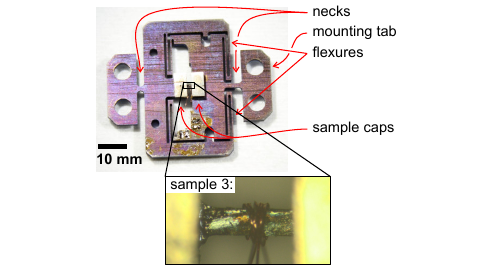}
\caption{\label{fig_sampleCarrier}Photograph of the sample carrier used to reduce hysteresis for samples 2 and 3. A photograph of sample 3 is also shown.}
\end{figure}

To cancel differential thermal contraction between the piezoelectric actuators and the body of the uniaxial stress cell, there are three actuators.  The outer two are joined
electrically, and ideally move identically. If they do not, a torque is generated within the cell that deforms the cell body and can yield a transverse displacement applied
across the sample.  That is a problem for measurement of \SRO{} under $\langle 110 \rangle$ uniaxial stress: a transverse displacement across the sample generates shear strain with
$\langle 100 \rangle$ principal axes within the sample, and $T_c$ of \SRO{} responds very sensitively to $\langle 100 \rangle$ shear strain.  Hysteresis in this torque, from
hysteresis in the actuator motion, resulted for sample~1 in hysteresis in $T_c$ that was large compared with the signal we aimed to measure.  To reduce this hysteresis, the
sample carrier photographed in Fig.~\ref{fig_sampleCarrier} was used for samples 2 and 3. It incorporates necks that attenuate the transmission of transverse displacements to the
sample.

\section{Calibration of the ECE} 
\label{Calib_ECE}

The elastocaloric effect $\eta \equiv d T/d \varepsilon$ is obtained by applying a small ac strain to a sample and measuring the ac temperature response. 
This ac strain can be superimposed onto much larger dc strains to measure the elastocaloric effect at different applied pressures. Details can be found in Ref.~\cite{Ikeda19_RSI, Li22_Nature}. The temperature response is measured with a Au/AuFe thermocouple,
which is glued by epoxy (Dupont 6838) to the center of the sample. 

\begin{figure}[ptb]
\includegraphics[width=87mm]{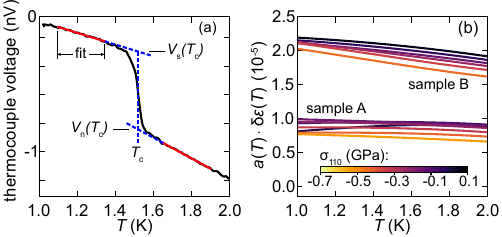}
\caption{\label{fig_ECEanalysis}(a) Extraction of $T_c$ from elastocaloric data. The data here are from sample A at 0~GPa. (b) Derivation of $A(T) \delta \varepsilon(T)$ for samples A and B, to obtain agreement with Eq.~\ref{eq_normalStateECE} in the normal state.}
\end{figure}

An example of extraction of $\Delta \eta / \eta_n$ is shown in Fig.~\ref{fig_ECEanalysis}(a). 
The measured thermocouple voltage is extrapolated from above and below the transition into the transition region. The temperature at which the data pass through the median line between these extrapolations is identified as $T_c$. The extrapolation from below to $T_c$ is identified as $V_\text{s}(T_c)$, and the extrapolation from above as $V_n(T_c)$, and $\Delta \eta / \eta_n$ is set to $(V_\text{s} - V_n)/V_n$.
As discussed above, applying
Eq.~\ref{eq_jump_ECE} with $dT_c/d\varepsilon$ set to 12.0~K yields $\gamma_1 = 0.43$~J/mol-K$^2$.

The conversion from thermocouple voltage $V$ to $\eta$ is given by
\[
\eta(T) = \frac{V(T)}{S(T) A(T) \delta \varepsilon (T)},
\]
where $S(T)$ is the Seebeck coefficient of the thermocouple, $0 < A < 1$ is the degree of adiabaticity, and $\delta \varepsilon$ is the strain oscillation amplitude.
$S(T)$ was determined by a thermocouple calibration in reference to a calibrated RuO$_2$ thermometer in Ref.~\cite{Li20_RSI}~\footnote{In the temperature range 0.15-4.5~K, the
thermopower is described by $S(T) = [10.1483\cdot\text{T/K} - 8.75772 \cdot \text{(T/K)}^2 + 4.00231 \cdot \text{(T/K)}^3 - 0.838741 \cdot \text{(T/K)}^4 + 0.0667604 \cdot
\text{(T/K)}^5] \mu\text{V/K}$}. We then derive the form of $A(T) \times \delta \varepsilon(T)$ that yields agreement in the normal state with Eq.~\ref{eq_normalStateECE}.  $A(T)
\times \delta \varepsilon(T)$ is obtained as a third-order polynomial, $\alpha_0 + \alpha_2 T^2 + \alpha_3 T^3$; the $T$-linear term is omitted because linear variation in the $T
\rightarrow 0$ limit is not generally expected. Results are shown in Fig.~\ref{fig_ECEanalysis}(b).  These derived curves are then extrapolated to below $T_c$, yielding
results shown in Fig.~\ref{fig_ECEData}.

\section{Jump in ECE } 
\label{Jump_ECE}



In Section~\ref{ECE} of the main text we showed how the elastocaloric coefficient for a single-component order parameter is derived from a Ginzburg-Landau ansatz.
In this appendix we derive the elastocaloric coefficient of two component order.
Eq.~\ref{eq_jump_ECE} can be rewritten to
\begin{equation}
\eta = \eta_n + \left(-\eta_n + \frac{dT_\text{c0}}{d\varepsilon}\right)\left(1 + \frac{T_\text{c0}}{\Delta C_0} \frac{C_n}{T}\right)^{-1}.
\label{eq_singleComponentECE}
\end{equation}

In the case of strain-split transitions,Eq.~\ref{eq_singleComponentECE} still applies at the first transition, at $T_c$. We have only a re-labelling of quantities, following above-described notation for split transitions. Let $\eta_1$ be the elastocaloric coefficient for the temperature range $T_2 < T < T_c$. 
We have:
\begin{equation}
\eta_1 = \eta_n + \left(-\eta_n + \frac{dT_c}{d\varepsilon}\right)\left(1 + \frac{T_{c}}{\Delta C_c} \frac{C_n}{T}\right)^{-1}.
\label{eq_twoComponentECE_eta1}
\end{equation}
Let $\eta_2$ be the elastocaloric coefficient at $T < T_2$. To obtain $\eta_2$, we take Eq.~\ref{eq_twoComponentECE_eta1} and replace $\eta_n$ with $\eta_1$, $\Delta C_c$ with $\Delta C_2$, and $T_c$ with $T_2$. 
We assume that $T_c$ and $T_2$ are both in the strain-linear regime, and apply the condition Eq.~\ref{eq_slopeCondition}. After simplification,
the result is:
\begin{equation}
\eta_2 = \eta_n + \left(-\eta_n + \frac{dT_{c0}}{d\varepsilon}\right) \left(1 + \frac{T_{c0}}{\Delta C_0} \frac{C_n}{T} \right)^{-1}.
\label{eq_twoComponentECE_eta2}
\end{equation}
If $\varepsilon$ is $\varepsilon_6$, then Eq.~\ref{eq_twoComponentECE_eta2} simplifies to
\begin{equation}
\eta_2 = \eta_n -\eta_n  \left(1 + \frac{T_{c0}}{\Delta C_0} \frac{C_n}{T} \right)^{-1}.
\label{eq_twoComponentECE_eta2b}
\end{equation}
Eq.~\ref{eq_twoComponentECE_eta2} is precisely the same as Eq.~\ref{eq_singleComponentECE}, though with a slight difference in interpretation: in
Eqs.~\ref{eq_twoComponentECE_eta2} and~\ref{eq_twoComponentECE_eta2b}, $T_\text{c0}$ is the transition temperature that \textit{would} be obtained in the absence of splitting. 

To evaluate the above expressions and obtain the left-hand bounding line in Fig.~\ref{fig_ECEExpectations}, we set $\gamma_0 = 0.038$~J/mol-K$^2$, $T_\text{c0} = 1.5$~K, $\gamma_1 = 0.43$~J/mol-K$^2$, $\beta = 0.000197$~J/mol-K$^4$, and $\varepsilon_{110} = -5 \cdot 10^{-4}$. 
Because the relevant data were taken almost entirely under compressive stress, we set $dT_c/d\varepsilon_{110} = 12.0\;\text{K} - \alpha_6 \times dT_c/d\varepsilon_6$.

\section{Component analysis} 
\label{Comp_appendix}

As a cross-check we can calculate from the stress dependence of $T_c$ under [110] pressure, $dT_c/d\sigma_{110} = - 64 \pm 7$~mK/GPa, and the stress dependence of
$T_c$ under [001] pressure, $dT_c/d\sigma_{001} = - 76 \pm 5$~mK/GPa \cite{Jerzembeck22_NatComm}, the stress dependence of $T_c$ under hydrostatic pressure and
compare this to the experimental results. Based on the 4~K elastic moduli reported in \cite{Ghosh21_NatPhys}, we obtain the following relations between strain and applied stress:
\begin{align}
\begin{aligned}
& \frac{d\varepsilon_d}{d\sigma_{110}} = 0.00307\;\text{GPa$^{-1}$}, \frac{d\varepsilon_3}{d\sigma_{110}} = -0.00102\;\text{GPa$^{-1}$}, \\
& \frac{d\varepsilon_d}{d\sigma_{001}} = -0.00204\;\text{GPa$^{-1}$}, \frac{d\varepsilon_3}{d\sigma_{001}} = 0.00457\;\text{GPa$^{-1}$}, \\
& \frac{d\varepsilon_d}{d\sigma_\text{hyd}} = 0.00411\;\text{GPa$^{-1}$}, \frac{d\varepsilon_3}{d\sigma_\text{hyd}} = 0.00254\;\text{GPa$^{-1}$}. 
\end{aligned}
\label{eq_stressStrainRelations}
\end{align}
In these expressions, $\varepsilon_d = \varepsilon_1 + \varepsilon_2$. So we have
\begin{align*}
\begin{aligned}
0.064 \pm 0.007\;\text{K} = 0.00307\frac{dT_c}{d\varepsilon_d} - 0.00102\frac{dT_c}{d\varepsilon_3},\\
0.076 \pm 0.005\;\text{K} = - 0.00204\frac{dT_c}{d\varepsilon_d} + 0.00457\frac{dT_c}{d\varepsilon_3}.
\end{aligned} 
\end{align*}
Solving these equations yields:
\begin{equation*}
\frac{dT_c}{d\varepsilon_d} = 31.0\pm 2.6~\text{K}, \frac{dT_c}{d\varepsilon_3} = 30.4\pm 1.8~\text{K}
\end{equation*}
Using the third line of Eq.~\eqref{eq_stressStrainRelations} yields $dT_c/d\sigma_\text{hyd} = 0.204\pm 0.012~\text{K/GPa}$.

\section{Ginzburg-Landau analysis} 
\label{sec_GL_appendix}

Here we analyze the response of a two-component order parameter $\Delta = (\Delta_{1}, \Delta_{2})^{\intercal}$ to $\sigma_6$ shear stress within the Ginzburg-Landau framework, under the assumptions of homogeneous strain and superconductivity.
While the analysis of a symmetry-protected two-component order parameter had already been done for the $D_{4h}$ point group~\cite{Benhabib21_NatPhys, Ghosh21_NatPhys}, the case of accidental degeneracy has not been analyzed in the literature to the degree of detail required for our analysis.
The symmetry-protected case corresponds to the two-dimensional irreducible representations E\textsubscript{g} and E\textsubscript{u} whose wavefunctions we may write as ($d_{xz}$, $d_{yz}$) and ($p_x$, $p_y$), respectively.
Accidental degeneracy could, in principle, be between any pair of one-dimensional irreducible representations (irreps).
We consider only those degenerate pairs that couple linearly to $\sigma_6$, which are A\textsubscript{1g} $\oplus$ B\textsubscript{2g} ($s$, $d_{xy}$) and B\textsubscript{1g} $\oplus$ A\textsubscript{2g} ($d_{x^2-y^2}$, $g_{xy(x^2-y^2)}$).
Odd-parity 1D irrep pairs, such as A\textsubscript{1u} $\oplus$ B\textsubscript{2u} and B\textsubscript{1u} $\oplus$ A\textsubscript{2u}, are also in principle possible, but not deemed likely due to Pauli limiting~\cite{Yonezawa13_PRL, Kinjo22_Science} and NMR Knight shift experiments~\cite{Pustogow19_Nature, Ishida20_JPSJ, Petsch20_PRL, Chronister21_PNAS, Jerzembeck23_PRB}.
Quadratic coupling to $\sigma_6$  does not induce a jump in the shear elastic modulus $c_{66}$ nor does it split the transition.

Before we proceed with the Ginzburg-Landau analysis, let us briefly discuss how we obtained the phase diagrams shown in Fig.~\ref{fig_schematicPhaseDiagrams}.
Introduce the bilinear forms
\begin{equation}
\phi_{\mu} = \Delta^{\dag} \tau_{\mu} \Delta, \label{eq_phimu-def}
\end{equation}
where $\tau_0$ is the $2 \times 2$ unit matrix and $\tau_{x, y, z}$ are Pauli matrices.
The transformation properties of $\phi_{\mu}$ are summarized in Table~\ref{tab:phi-irreps}.
A sufficient condition for a cusp in $T_c(\sigma_6)$ is that there exists a $\phi_{\mu}$ that transforms like the shear strain $\sigma_6 \in$ B$_{\text{2g}}^{+}$.
In our case, this is only possible for $\phi_x$.
If $\phi_x$ acquires a nonzero expectation value below $T_c$ at $\varepsilon_6 = 0$, then strain acts like a conjugate field that lifts the degeneracy between $\pm \langle{\phi_x\rangle}$, and only one transition takes place since the symmetry associated with $\phi_x$ is already broken.
If, on the other hand,  $\phi_x$ is not the bilinear that acquires a finite expectation value below $T_c$ at $\varepsilon_6 = 0$, an additional symmetry can still break, resulting in a second transition.

\begin{table}
\hfill {\renewcommand{\arraystretch}{1.3}
\renewcommand{\tabcolsep}{10pt}
\begin{tabular}{|c|c|} \hline\hline
\multicolumn{2}{|c|}{$\Delta \in$ E\textsubscript{g,u}} \tabularnewline 
bilinear & irrep \tabularnewline \hline
$\phi_{0}$ & A$_{\text{1g}}^{+}$ \tabularnewline 
$\phi_{x}$ & B$_{\text{2g}}^{+}$ \tabularnewline 
$\phi_{y}$ & A$_{\text{2g}}^{-}$ \tabularnewline 
$\phi_{z}$ & B$_{\text{1g}}^{+}$ \tabularnewline[1pt] \hline\hline
\end{tabular}} \hfill %
{\renewcommand{\arraystretch}{1.3}
\renewcommand{\tabcolsep}{10pt}
\begin{tabular}{|c|c|} \hline\hline
\multicolumn{2}{|c|}{$\Delta_1 \in \Gamma_{1}$, $\Delta_2 \in \Gamma_{2}$} \tabularnewline
bilinear & irrep \tabularnewline \hline
$\phi_{0}$ & A$_{\text{1g}}^{+}$ \tabularnewline 
$\phi_{x}$ & $(\Gamma_{1} \otimes \Gamma_{2})^{+}$ \tabularnewline 
$\phi_{y}$ & $(\Gamma_{1} \otimes \Gamma_{2})^{-}$ \tabularnewline 
$\phi_{z}$ & A$_{\text{1g}}^{+}$ \tabularnewline[1pt] \hline\hline
\end{tabular}} \hfill \hphantom{a} \\
\caption{Irreducible representations (irreps) of $D_{4h}$ under which the bilinear forms $\phi_{\mu}$ transform.
The bilinears are constructed from two-component order parameters $\Delta = (\Delta_{1}, \Delta_{2})$ according to Eq.~\eqref{eq_phimu-def}.
$\Delta$ belongs to the 2D irreps $E_{g,u}$ on the left, whereas on the right $\Delta_{1,2}$ belong to two 1D irreps  $\Gamma_{1, 2}$, respectively.
The $\pm$ irrep superscripts describe the behavior under time reversal.
Only accidentally degenerate pairs whose $\Gamma_{1} \otimes \Gamma_{2} =$ B\textsubscript{2g} are analyzed in this appendix.}
\label{tab:phi-irreps}
\end{table}

The Ginzburg-Landau expansion of the free energy in the absence of stress is given by
\begin{align}
\begin{aligned}
F = F_n &+ \frac{a}{2} \phi_0 + \frac{u}{4} \phi_0^2 + \sum_{\mu=x,y,z} \frac{v_{\mu}}{4} \phi_{\mu}^2 \\
&\qquad + \frac{\tilde{a}}{2} \phi_z + \frac{\tilde{v}}{4} \phi_{0} \phi_{z}.
\end{aligned} \label{eq_GLexp}
\end{align}
From Table~\ref{tab:phi-irreps}, it is easy to see that this is the most general form of an invariant (A$_{\text{1g}}^{+}$) function that is quadratic in $\phi_{\mu}$, that is quartic in $\Delta$.
Due to the Fierz identity
\begin{equation}
\phi_0^2 = \sum_{\mu=x,y,z} \phi_{\mu}^2,
\end{equation}
there is a redundancy between the $v_{\mu} \phi_{\mu}^2$ terms that we eliminate by setting
\begin{equation}
v_z = 0.
\end{equation}

In the case of a symmetry-protected degeneracy, $\phi_{z}$ transforms under B\textsubscript{1g} and therefore $\tilde{a} = \tilde{v} = 0$.
Below the transition temperature $T_\text{c0}$, the quadratic coefficient changes sign.
To leading order in temperature, $a(T)$ is thus linear in $T$ with a positive slope $\dot{a} > 0$:
\begin{equation}
a = (T - T_\text{c0}) \times \dot{a}, \label{eq_aofT}
\end{equation}
whereas the quartic coefficients are $T$-independent.

When $\Delta_{1, 2}$ belong to two 1D irreps, $\phi_{z}$ transforms trivially and both $\tilde{a}$ and $\tilde{v}$ are allowed to be finite.
However, since $\Delta_1$ and $\Delta_2$ are unrelated by symmetry, we may rescale them $(\Delta_1, \Delta_2) \mapsto (s\Delta_1, s^{-1}\Delta_2)$ by a factor $s = \big(\tfrac{u + \tilde{v}}{u - \tilde{v}}\big)^{1/8}$ so that after the rescaling
\begin{align}
\tilde{v} = 0,
\end{align}
which we assume henceforth.
Regarding $\tilde{a}$, in the expansion $F = \dot{a}_1 \left(T-T_{\text{c0},1}\right) \left|\Delta_{1}\right|^{2} + \dot{a}_2 \left(T-T_{\text{c0},2}\right) \left|\Delta_{2}\right|^{2} + \cdots$ the fine-tuning of the two transition temperatures corresponds to the requirement that $T_{\text{c0},1} = T_{\text{c0},2} \equiv T_\text{c0}$.
Hence $a(T)$ is given by Eq.~\eqref{eq_aofT} with $\dot{a} = \dot{a}_1 + \dot{a}_2$, while
\begin{equation}
\tilde{a}(T) = \alpha \times a(T)
\end{equation}
for a $T$-independent coefficient $\alpha = \tfrac{\dot{a}_1 - \dot{a}_2}{\dot{a}_1 + \dot{a}_2}$. $\alpha$ can take any value in between $-1$ and $1$ and reflects the absence of a symmetry transformation connecting $\Delta_1$ and $\Delta_2$.
Thus in the symmetry-protected case the only formal difference is that $\alpha = 0$, given that $\tilde{v} = 0$ in both cases.

Let us now include elasticity.
When strains $\varepsilon_i$ are present in the system, they couple to the superconductivity via
\begin{equation}
F_{c} = \sum\limits_{i=1}^6 \sum\limits_{a,b=1}^2 \lambda_{iab} \varepsilon_i \Delta_a^* \Delta_b,
\end{equation}
where $\lambda_{iab}$ are the coupling constants.
As it turn out, when the elastic free energy is quadratic in $\varepsilon_i$,
\begin{equation}
F_{\varepsilon} = \frac{1}{2}\sum\limits_{i,j=1}^6 c_{ij,0} \varepsilon_i \varepsilon_j,
\end{equation}
one may decouple the elastic and superconducting parts of the free energy, greatly simplifying the free energy minimization problem.
Here $c_{ij,0}$ is the elastic tensor in the absence of superconductivity.
This decoupling is accomplished by introducing the ``external'' strain
\begin{equation}
\varepsilon_{i,0} \equiv \varepsilon_i + \sum\limits_{j=1}^6 \sum\limits_{a,b=1}^2 c_{ij,0}^{-1} \lambda_{iab} \Delta_a^* \Delta_b, \label{eq_externalStrain}
\end{equation}
which is decoupled from $\Delta$ and directly related to the external stress: $\varepsilon_{i,0} = \sum_{j=1}^6 c_{ij,0}^{-1} \sigma_j$.
It is the strain that would be obtained under a given set of stresses in the absence of superconductivity.

In practice, the difference between $\varepsilon_{i,0}$ and the total strain $\varepsilon_i$ is negligible for \SRO: the larger of the two reported values of $\Delta c_{66}$ is $\sim 10^{-5} c_{66,0}$~\cite{Benhabib21_NatPhys, Ghosh21_NatPhys}, and the experimental upper limit on any spontaneous nematic strain is on the order of $10^{-8}$ [Eq.~\eqref{eq_nematicUpperLimit}], far smaller than the scale of the strains applied in this work.
For this reason, in the main text we make no distinction between $\varepsilon_{i,0}$ and $\varepsilon_i$.
Here, we retain this distinction to be able to calculate the jump in the shear modulus $\Delta c_{66}$.

In the presence of $\sigma_6$ external shear stress, the total free energy after decoupling therefore equals
\begin{equation}
F = F_n + F_{\varepsilon0} + F_{\Delta0},
\end{equation}
where the elastic part is
\begin{equation}
F_{\varepsilon0} = \frac{1}{2} c_{66,0} \varepsilon_{6,0}^2 - \sigma_6 \varepsilon_{6,0},
\end{equation}
and the superconducting part is
\begin{equation}
\begin{aligned}
F_{\Delta0} &= \frac{a}{2} \big(|\Delta_1|^2 + |\Delta_2|^2\big) + \alpha \frac{a}{2} \big(|\Delta_1|^2 - |\Delta_2|^2\big) \\
&\quad + \frac{u}{4} \big(|\Delta_1|^2 + |\Delta_2|^2\big)^2 + \gamma u |\Delta_1|^2 |\Delta_2|^2 \\
&\quad + \gamma' \frac{u}{2} \big(\Delta_1^{*} \Delta_1^{*} \Delta_2 \Delta_2 + \Delta_2^{*} \Delta_2^{*} \Delta_1 \Delta_1\big) \\
&\quad + \sigma_6 c_{66,0}^{-1} \lambda_6 \big(\Delta_1^{*} \Delta_2 + \Delta_2^{*} \Delta_1\big).
\end{aligned} \label{eq_F_Delta}
\end{equation}
Here we have explicitly written out $F_{\Delta0}$ in terms of $\Delta$ instead of $\phi_{\mu}$.
The form of the coupling to $\sigma_6 \in$ B$_{\text{2g}}^{+}$ follows from Table~\ref{tab:phi-irreps}; for the accidentally degenerate case, we assumed $\Gamma_{1} \otimes \Gamma_{2} =$ B\textsubscript{2g}.
By enacting $(\Delta_1, \Delta_2) \mapsto (\Delta_1, -\Delta_2)$, we can always make $\lambda_6 > 0$, which we henceforth assume.
For later convenience, we parameterized the quartic coefficients in terms of $\gamma, \gamma'$ which are related to the $v_{x, y}$ of Eq.~\eqref{eq_GLexp} via $v_x = (\gamma + \gamma') u$ and $v_y = (\gamma - \gamma') u$.
In shifting from $\varepsilon_i$ to $\varepsilon_{i,0}$, the quartic coefficients $u, \gamma, \gamma'$ have been renormalized.

The free energy is bounded from below when $u > 0$ and $\gamma - |\gamma^\prime| > -1$; these constraints define the physical part of the parameter space.
To find the minimum of $F_{\Delta0}$, we use the spherical parameterization
\begin{align}
\begin{pmatrix}
\Delta_1 \\
\Delta_2
\end{pmatrix} &= \Delta_0 \begin{pmatrix}
\cos \theta \\
\sin \theta \, e^{i \phi}
\end{pmatrix}
\end{align}
in terms of which
\begin{equation}
F_{\Delta0} = A(\theta, \phi) \frac{a}{2} \Delta_0^2 + U(\theta, \phi) \frac{u}{4} \Delta_0^4,
\label{eq_F_reduced}
\end{equation}
where
\begin{align}
A(\theta, \phi) &= 1 + \alpha \cos(2\theta) + \beta \sin(2\theta)\cos(\phi), \label{eq_A} \\
U(\theta, \phi) &= 1 + \Gamma(\phi) \sin^2(2\theta), \\
\Gamma(\phi) &= \gamma + \gamma^\prime\cos(2\phi).
\end{align}
Here we have introduced the shorthand:
\begin{equation}
\beta \equiv \frac{2\lambda_6\varepsilon_{6,0}}{a}.
\end{equation}
The saddle point equations for the nontrivial solution whose $\Delta_0^2 = - a A / (u U) > 0$ are:
\begin{align}
0 & = \sin(\phi) \sin(2\theta) \left[\gamma^\prime \cos(\phi) \sin(2\theta) - \frac{\beta U}{2 A}\right], \label{eq_saddlePoint1} \\
0 & = \Gamma \sin( 2\theta) \cos(2\theta) + \frac{U}{A}\big[\alpha \sin (2\theta) - \beta \cos(\phi) \cos(2 \theta)\big]. \label{eq_saddlePoint2}
\end{align}

\subsection{Solutions for \boldmath{$\sigma_6 = 0$}}
In the absence of applied stress ($\beta = 0$), these saddle point equations are easily solved.
They give three classes of solutions.
\begin{itemize}
\item $\Delta_1$ or $\Delta_2$ only: $\theta = 0$ or $\pi/2$, $\mathbf{\Delta} \sim (1,0)$ or $(0,1)$.
\item B\textsubscript{2g}-nematic: $\theta = \frac{1}{2}\arccos (-\alpha x_+)$, $\phi = 0$ or $\pi$, \\ $\mathbf{\Delta} \sim (1, \pm 1)$.
\item TRSB: $\theta = \frac{1}{2}\arccos (- \alpha x_-)$, $\phi = \pm \frac{\pi}{2}$, $\mathbf{\Delta} \sim (1, \pm i)$.
\end{itemize}
In these expressions
\begin{equation}
x_{\pm} \equiv \frac{1 + \gamma \pm \gamma'}{\gamma \pm \gamma'}.
\end{equation}
In the case of symmetry-protected degeneracy ($\alpha = 0$), the $\Delta_1$ only and $\Delta_2$ only ground states are degenerate and may be identified with B\textsubscript{1g}-nematic order.

The free energy values for these solutions are:
\begin{align}
F_{\Delta0} &= - \frac{a^2}{4 u} \times \begin{cases}
(1 + \alpha)^2, & \text{$\Delta_1$ only} \\
(1 - \alpha)^2, & \text{$\Delta_2$ only} \\
\displaystyle \frac{1 - \alpha^2 x_{+}}{1 + \gamma + \gamma^\prime}, & \text{B\textsubscript{2g}-nem.} \\
\displaystyle \frac{1 - \alpha^2 x_{-}}{1 + \gamma - \gamma^\prime}, & \text{TRSB}
\end{cases} \label{eq_app-free-energy}
\end{align}
The preferred global minimum is:
\begin{align}
\begin{aligned}
\gamma - |\gamma^\prime| > -\frac{|\alpha|}{1 + |\alpha|},\;\; & \alpha > 0:\;\; && \text{$\Delta_1$ only} \\
\gamma - |\gamma^\prime| > -\frac{|\alpha|}{1 + |\alpha|},\;\; & \alpha < 0:\;\; && \text{$\Delta_2$ only} \\
-1 < \gamma + \gamma^\prime < -\frac{|\alpha|}{1 + |\alpha|},\;\; & \gamma^\prime < 0:\;\; && \text{B\textsubscript{2g}-nem.} \\
-1 < \gamma - \gamma^\prime < -\frac{|\alpha|}{1 + |\alpha|},\;\; & \gamma^\prime > 0:\;\; && \text{TRSB}
\end{aligned} \label{eq_paramRegions}
\end{align}
For the $\alpha = 0$ case, the corresponding phase diagram is shown in Fig.~\ref{fig_triplePointTuning} of the main text.
For finite $\alpha$, the B\textsubscript{1g}-nematic region of Fig.~\ref{fig_triplePointTuning} becomes $\Delta_1$ or $\Delta_2$ only, and its lower edge $\gamma - |\gamma^\prime| = 0$ is shifted downward by $\tfrac{|\alpha|}{1 + |\alpha|}$.

\subsection{Solutions for \boldmath{$\sigma_6 \neq 0$}}
First consider $T > T_\text{c0}$.
In this case, $F_{\Delta0} < 0$ is only obtained when $A(\theta, \phi) < 0$.
By minimizing Eq.~\eqref{eq_A}, we see that the minimum of $A(\theta, \phi)$ is $1 - \sqrt{\alpha^2 + \beta^2}$ and has $\phi = 0$ or $\pi$ with $\theta \neq 0$, which corresponds to B\textsubscript{2g}-nematic order.
Hence
\begin{equation}
T_c = T_\text{c0} + \frac{\lambda_6 |\varepsilon_{6,0}|}{\dot{a}} \frac{2}{\sqrt{1-\alpha^2}}
\end{equation}
and the symmetry is B\textsubscript{2g}-nematic.

Now consider reducing $T$ below $T_c$.
As illustrated in Fig.~\ref{fig_schematicPhaseDiagrams}, a second transition takes place when the ground state breaks time-reversal symmetry, whether the degeneracy is symmetry-protected or not, and when the ground state is B\textsubscript{1g}-nematic.
In the latter case, the degeneracy must be symmetry-protected because only then is the $(\Delta_1, \Delta_2) \mapsto (\Delta_2, \Delta_1)$ symmetry present which forbids a smooth crossover between B\textsubscript{1g} and B\textsubscript{2g}-nematic states.

To determine the lower transition temperature $T_2$, we need to solve the saddle point equations~(\ref{eq_saddlePoint1},\ref{eq_saddlePoint2}) and figure out which solution yields the smallest free energy.
For the nematic case, $\phi = 0$ or $\pi$ and $\theta$ is determined by the transcendental equation
\begin{equation}
\left(\cos(2\theta) + \alpha x_+\right)\sin(2\theta) = \frac{\beta}{\gamma + \gamma^\prime}\cos(2\theta). \label{eq_nematicSaddlePointEq}
\end{equation}
When the $\sigma_6 = \beta = 0$ ground state is B\textsubscript{2g}-nematic, $\theta$ of the global minimum changes smoothly with $\sigma_6$ and there is no second transition.
The same happens when $\Delta_1$ or $\Delta_2$ are the ground states and $\alpha \neq 0$: we have a smooth crossover, as one can show by analyzing the bifurcation of the solutions.

When the ground state is B\textsubscript{1g}-nematic and $\alpha = 0$, B\textsubscript{1g}-nematic solutions overtake the B\textsubscript{2g}-nematic solutions below $|\beta| = \gamma + \gamma'$, yielding
\begin{equation}
T_2 = T_\text{c0} - \frac{\lambda_6 |\varepsilon_{6,0}|}{\dot{a}} \frac{2}{\gamma + \gamma'}.
\end{equation}

When the ground state is TRSB with symmetry-protected degeneracy:
\begin{equation}
T_2 = T_\text{c0} - \frac{\lambda_6 |\varepsilon_{6,0}|}{\dot{a}} \frac{1 + \gamma - \gamma'}{\gamma'}. \label{eq_T2_symmProtectedTRSB}
\end{equation}
Along the line $\gamma = \gamma'$ that is the boundary between the B\textsubscript{1g} and TRSB regions of the $\alpha=0$ parameter space [Eq.~\eqref{eq_paramRegions}], these two expressions for $T_2$ agree.
When the ground state is TRSB with accidental degeneracy:
\begin{equation}
T_2 = T_\text{c0} - \frac{\lambda_6 |\varepsilon_{6,0}|}{\dot{a}} \frac{1 + \gamma - \gamma'}{\gamma' \sqrt{1 - \alpha^2 x_-^2}}. \label{eq_T2_accidentalTRSB}
\end{equation}
In the TRSB case, one may solve the saddle point equations in closed-form:
\begin{align}
\theta & = \frac{1}{2}\arccos(- \alpha x_-), \\
\phi & = \pm \arccos\left(\frac{\lambda \varepsilon_{6,0}}{a} \frac{1 + \gamma - \gamma^\prime}{\gamma' \sqrt{1 - \alpha^2 x_-^2}}\right), \\
F_{\Delta0} & = -\frac{a^2}{4 u} \frac{1 - \alpha^2 x_{-}}{1 + \gamma - \gamma^\prime} - \frac{\lambda_6^2 \varepsilon_{6,0}^2}{2u\gamma^\prime}. \label{eq_TRSBFreeEnergy}
\end{align}

\subsection{Ehrenfest relations} \label{sec_Ehrenfest_relations_appendix}
The jump in the heat capacity across the superconducting transition is given by:
\begin{equation}
\frac{\Delta C_0}{T_\text{c0}} = - \left.\frac{\partial^2 F_{\Delta0}}{\partial T^2}\right|_{T = T_\text{c0}, \sigma_6 = 0}.
\end{equation}
From the free energy expressions of Eq.~\eqref{eq_app-free-energy}:
\begin{align}
\frac{\Delta C_0}{T_\text{c0}} &= \frac{\dot{a}^2}{2u} \times \begin{cases}
(1 + |\alpha|)^2, & \text{$\Delta_1$ or $\Delta_2$ only} \\
\displaystyle \frac{1 - \alpha^2 x_{+}}{1 + \gamma + \gamma^\prime}, & \text{B\textsubscript{2g}-nem.} \\
\displaystyle \frac{1 - \alpha^2 x_{-}}{1 + \gamma - \gamma^\prime}, & \text{TRSB}
\end{cases}
\end{align}

The shear elastic modulus $c_{66}$ below $T_c$ is given by
\begin{equation}
\frac{1}{c_{66}} = \frac{1}{c_{66,0}} - \left.\frac{\partial^2 F_{\Delta0}}{\partial \sigma_6^2}\right|_{T, \sigma_6 = 0}.
\end{equation}
The jump $\Delta c_{66} = c_{66,0} - c_{66}|_{T=T_\text{c0}}$ is the difference between $c_{66}$ just above $T_\text{c0}$ and that just below it.

When the ground state is $\Delta_1$ or $\Delta_2$ only,
\begin{equation}
\Delta c_{66} = \frac{2\lambda_6^2}{u} \frac{1+|\alpha|}{(\gamma + \gamma') (1 + |\alpha| x_{+})}.
\end{equation}
This is derived by solving Eq.~\eqref{eq_nematicSaddlePointEq} for small $\beta$.
In the case of symmetry-enforced degeneracy ($\alpha = 0$), that is, B\textsubscript{1g}-nematic ground states, one obtains the following Ehrenfest relation:
\begin{align}
\Delta c_{66} &= \frac{\Delta C_0}{T_\text{c0}} \left|\frac{dT_c}{d\varepsilon_{6,0}}\right| \left|\frac{dT_2}{d\varepsilon_{6,0}}\right|.
\end{align}
In the general $\alpha \neq 0$ case, we could try using $T_c$ instead of $T_2$ above, but the corresponding dimenionless ratio can be any positive real number, depending on the values of $\alpha$ and $\gamma + \gamma'$ which we do not know.

When the ground state is B\textsubscript{2g}-nematic, by solving Eq.~\eqref{eq_nematicSaddlePointEq} one finds that
\begin{equation}
\Delta c_{66} = \frac{2\lambda_6^2}{u} \frac{1 - \alpha^2 x_+^3}{(1 + \gamma + \gamma') (1 - \alpha^2 x_{+}^2)},
\end{equation}
and therefore
\begin{equation}
\frac{\Delta c_{66}}{\displaystyle \frac{\Delta C_0}{T_\text{c0}} \left|\frac{dT_c}{d\varepsilon_{6,0}}\right| \left|\frac{dT_c}{d\varepsilon_{6,0}}\right|} = \frac{(1 - \alpha^2) (1 - \alpha^2 x_+^3)}{(1 - \alpha^2 x_+) (1 - \alpha^2 x_+^2)}.
\end{equation}
When $\alpha = 0$, this expression reduces to the standard Ehrenfest relation.
The stability condition for B\textsubscript{2g}-nematic order [Eq.~\eqref{eq_paramRegions}] corresponds to $-1/|\alpha| < x_+ < 0$ and the right-hand side can equal any number between $(1 - \alpha^2)$ and $+ \infty$ for $\alpha \neq 0$ and $x_{+}$ in this range.

When the ground state is TRSB, the second derivative of Eq.~\eqref{eq_TRSBFreeEnergy} with respect to $\varepsilon_{6,0}$ yields
\begin{equation}
\Delta c_{66} = \frac{\lambda_6^2}{u \gamma^\prime}.
\end{equation}
The corresponding Ehrenfest relation takes the form:
\begin{equation}
\frac{\Delta c_{66}}{\displaystyle \frac{\Delta C_0}{T_\text{c0}} \left|\frac{dT_c}{d\varepsilon_{6,0}}\right| \left|\frac{dT_2}{d\varepsilon_{6,0}}\right|} = \frac{\sqrt{1-\alpha^2}\sqrt{1 -
\alpha^2 x_-^2}}{1 - \alpha^2 x_-} \leq 1.
\label{eq_fullEhrenfestRelation}
\end{equation}
In the $-1/|\alpha| < x_- < 0$ region where TRSB is the ground state [Eq.~\eqref{eq_paramRegions}], the right-hand side takes values in between $0$ and $1$, and for $\alpha = 0$ equals $1$.

\subsection{Ratio relations} 
\label{sec_ratio_relations_appendix}
Here we show that the ratios of the jumps at the upper and lower transitions are related.
These relations hold only for the symmetry-protected case ($\alpha = 0$).
Denote $\Delta C_c$ and $\Delta c_{66,\text{c}}$ the jumps at the upper transition ($T = T_c$), and $\Delta C_2$ and $\Delta c_{66,2}$ the jumps at the lower transition ($T = T_2$).

The jumps at the upper transition are ($\alpha = 0$, $\sigma_6 \neq 0$):
\begin{align}
\frac{\Delta C_c}{T_c} &= \frac{\dot{a}^2}{2 u} \frac{1}{1 + \gamma + \gamma'}, \\
\Delta c_{66,\text{c}} &= \frac{2 \lambda_6^2}{u} \frac{1}{1 + \gamma + \gamma'}.
\end{align}
The jumps at the lower transition are ($\alpha = 0$, $\sigma_6 \neq 0$):
\begin{align}
\frac{\Delta C_2}{T_2} &= \frac{\dot{a}^2}{2 u} \times \begin{cases}
\displaystyle \frac{\gamma + \gamma'}{1 + \gamma + \gamma'}, & \text{B\textsubscript{1g}-nem.} \\
\displaystyle \frac{2 \gamma'}{(1 + \gamma + \gamma') (1 + \gamma - \gamma')}, & \text{TRSB}
\end{cases} \\
\Delta c_{66,2} &= \frac{2 \lambda_6^2}{u} \times \begin{cases}
\displaystyle \frac{1}{(\gamma + \gamma') (1 + \gamma + \gamma')}, & \text{B\textsubscript{1g}-nem.} \\[8pt]
\displaystyle \frac{1 + \gamma - \gamma'}{(1 + \gamma + \gamma') 2 \gamma'}, & \text{TRSB}
\end{cases}
\end{align}
To find these expressions, we had to solve Eq.~\eqref{eq_nematicSaddlePointEq} around the $\beta$ at which the solutions bifurcate.
Note that $\Delta C_c/T_c + \Delta C_2/T_2$ and $\Delta c_{66,\text{c}} + \Delta c_{66,2}$ reproduce the previous $\Delta C_0/T_\text{c0}$ and $\Delta c_{66}$ with $\alpha = 0$.
Combining, we obtain the ratio relations:
\begin{equation}
\begin{aligned}
\frac{\displaystyle \left|\frac{dT_2}{d\varepsilon_{6,0}}\right|}{\displaystyle \left|\frac{dT_c}{d\varepsilon_{6,0}}\right|} &= \frac{\displaystyle \frac{\Delta C_c}{T_c}}{\displaystyle \frac{\Delta C_2}{T_2}} = \frac{\Delta c_{66,2}}{\Delta c_{66,\text{c}}} \\
&= \begin{cases}
\displaystyle \frac{1}{\gamma + \gamma'}, & \text{B\textsubscript{1g}-nem.} \\[8pt]
\displaystyle \frac{1 + \gamma - \gamma'}{2 \gamma'}, & \text{TRSB}
\end{cases}
\end{aligned} \label{eq_ratioRelationsAppendix}
\end{equation}

\subsection{Nematic strain}
The second term in Eq.~\eqref{eq_externalStrain} defines the ``internal'' strain, which is the strain generated by the superconducting order parameter:
\begin{align}
\begin{aligned}
\varepsilon_6^\text{nem} &= -\frac{\lambda_6}{c_{66}}\big(\Delta_1^* \Delta_2 + \Delta_2^* \Delta_1\big) \\
&= -\frac{\lambda_6}{c_{66}}\Delta_0^2 \sin (2\theta) \cos \phi.
\end{aligned}
\end{align}
Due to the proportionality to $\cos \phi$, when $\sigma_6 = 0$ only the B\textsubscript{2g}-nematic states generate a non-zero $\varepsilon_6$.
Its value is bounded from above through
\begin{equation}
\frac{c_{66,0} |\varepsilon_6^\text{nem}|}{\displaystyle \frac{\Delta C_0}{T_\text{c0}} \left|\frac{dT_c}{d\varepsilon_{6,0}}\right| |T - T_\text{c0}|} = \frac{\sqrt{1 - \alpha^2} \sqrt{1 - \alpha^2 x_+^2}}{1 - \alpha^2 x_+} \leq 1, \label{eq_nematicUpperLimit}
\end{equation}
where the right-hand side is in between $0$ and $1$ in the range $-1/|\alpha| < x_+ < 0$ where B\textsubscript{2g}-nematic order is preferred [Eq.~\eqref{eq_paramRegions}], and for $\alpha = 0$ equals $1$.

\subsection{Case of B\textsubscript{1g} stress} \label{sec_B1g_relations_appendix}
As discussed in the main text, if one combines the measurements of the current paper with those performed under $\langle 100 \rangle$ uniaxial stress~\cite{Li21_PNAS}, one can put tight constraints on where precisely \SRO{} must be in the phase diagram of Fig.~\ref{fig_triplePointTuning}.

To make contact with the measurements under $\langle 100 \rangle$ uniaxial stress, here we briefly summarize the results of the Ginzburg-Landau analysis for B\textsubscript{1g} stress, $\sigma_{\text{B\textsubscript{1g}}} = \tfrac{1}{2} (\sigma_1 - \sigma_2) = \sigma_{100} / 2$.
Superconductivity couples linearly to B\textsubscript{1g} stress only in the case of symmetry-protected degeneracy ($\alpha = 0$), which we henceforth consider.

The coupling to B\textsubscript{1g} stress takes the form
\begin{align}
F_{\Delta0} = \cdots + \sigma_{\text{B\textsubscript{1g}}} c_{\text{B\textsubscript{1g}}}^{-1} \lambda_{\text{B\textsubscript{1g}}} \big(|\Delta_1|^2 - |\Delta_2|^2\big),
\end{align}
where $c_{\text{B\textsubscript{1g}}} = \tfrac{1}{2} (c_{11} - c_{12})$.
By a rotation
\begin{equation}
\tilde{\Delta} = \frac{1}{\sqrt{2}} \begin{pmatrix}
1 & -1 \\
1 & 1
\end{pmatrix} \Delta
\end{equation}
and reparameterization
\begin{align}
\tilde{u} &= (1 + \gamma + \gamma') u, \\
\tilde{\gamma} &= \frac{-\tfrac{1}{2} \gamma - \tfrac{3}{2} \gamma'}{1 + \gamma + \gamma'}, \\
\tilde{\gamma}' &= \frac{-\tfrac{1}{2} \gamma + \tfrac{1}{2} \gamma'}{1 + \gamma + \gamma'},
\end{align}
one obtains a free energy identical in form to Eq.~\eqref{eq_F_Delta}.
Hence all the previous formulas carry over if we replace $u, \gamma, \gamma', \lambda_6$ with $\tilde{u}, \tilde{\gamma}, \tilde{\gamma'}, \lambda_{\text{B\textsubscript{1g}}}$, and exchanges what one identifies as B\textsubscript{1g} with B\textsubscript{2g}, and vice versa.

The upper transition temperature is given by:
\begin{align}
T_c &= T_\text{c0} + \frac{2 \lambda_{\text{B\textsubscript{1g}}} |\varepsilon_{\text{B\textsubscript{1g}},0}|}{\dot{a}}.
\end{align}
At finite B\textsubscript{1g} stress, the superconductivity is B\textsubscript{1g}-nematic slightly below $T_c$.
When B\textsubscript{1g}-nematic pairing is the ground state, there is no second transition.
For the other two cases:
\begin{align}
T_2 &= T_\text{c0} - \frac{2 \lambda_{\text{B\textsubscript{1g}}} |\varepsilon_{\text{B\textsubscript{1g}},0}|}{\dot{a}} \times \begin{cases}
- x_{+}, & \text{B\textsubscript{2g}-nem.} \\
- x_{-}, & \text{TRSB}
\end{cases}
\end{align}

The heat capacity jumps:
\begin{align}
\frac{\Delta C_c}{T_c} &= \frac{\dot{a}^2}{2 u}, \\
\frac{\Delta C_2}{T_2} &= \frac{\dot{a}^2}{2 u} \times \begin{cases}
-1/x_{+}, & \text{for B\textsubscript{2g}-nem.} \\
-1/x_{-}, & \text{for TRSB}
\end{cases}
\end{align}
The jumps in the B\textsubscript{1g} elastic constants:
\begin{align}
\Delta c_{\text{B\textsubscript{1g}},\text{c}} &= \frac{2 \lambda_{\text{B\textsubscript{1g}}}^2}{u}, \\
\Delta c_{\text{B\textsubscript{1g}},2} &= \frac{2 \lambda_{\text{B\textsubscript{1g}}}^2}{u} \times \begin{cases}
-x_{+}, & \text{for B\textsubscript{2g}-nem.} \\
-x_{-}, & \text{for TRSB}
\end{cases}
\end{align}
The total jumps are obtained by summing the jumps at the upper and lower transition, if it takes place.

The Ehrenfest relation for B\textsubscript{1g}-nematic states:
\begin{align}
\Delta c_{\text{B\textsubscript{1g}}} &= \frac{\Delta C_0}{T_\text{c0}} \left|\frac{dT_c}{d\varepsilon_{\text{B\textsubscript{1g}},0}}\right| \left|\frac{dT_c}{d\varepsilon_{\text{B\textsubscript{1g}},0}}\right|.
\end{align}
The Ehrenfest relation when B\textsubscript{2g}-nematic or TRSB pairing is preferred in the absence of stress:
\begin{align}
\Delta c_{\text{B\textsubscript{1g}}} &= \frac{\Delta C_0}{T_\text{c0}} \left|\frac{dT_c}{d\varepsilon_{\text{B\textsubscript{1g}},0}}\right| \left|\frac{dT_2}{d\varepsilon_{\text{B\textsubscript{1g}},0}}\right|.
\end{align}
Ratio relations:
\begin{equation}
\begin{aligned}
\frac{\displaystyle \left|\frac{dT_2}{d\varepsilon_{\text{B\textsubscript{1g}},0}}\right|}{\displaystyle \left|\frac{dT_c}{d\varepsilon_{\text{B\textsubscript{1g}},0}}\right|} &= \frac{\displaystyle \frac{\Delta C_c}{T_c}}{\displaystyle \frac{\Delta C_2}{T_2}} = \frac{\Delta c_{\text{B\textsubscript{1g}},2}}{\Delta c_{\text{B\textsubscript{1g}},\text{c}}} \\
&= \begin{cases}
-x_{+}, & \text{B\textsubscript{2g}-nem.} \\
-x_{-}, & \text{TRSB}
\end{cases}
\end{aligned} \label{eq_ratioRelationsAppendixB1g}
\end{equation}

\bibliography{bibliography.bib}
\end{document}